\documentclass[preprint,5p,times,twocolumn]{elsarticle}
\usepackage{amsmath}
\usepackage{amssymb}
\usepackage{verbatim}
\usepackage[colorlinks,linkcolor=blue,citecolor=blue,urlcolor=blue]{hyperref}
\usepackage{booktabs}
\usepackage[dvipsnames]{xcolor}
\usepackage{listings}
\usepackage{lipsum}
\usepackage{makecell}

\usepackage[nomain,toc,acronym,automake, nonumberlist, style=superheaderborder,nogroupskip]{glossaries} 	
\makeglossaries										

\usepackage{xcolor}
\definecolor{commentgreen}{RGB}{2,112,10}
\definecolor{eminence}{RGB}{108,48,130}
\definecolor{weborange}{RGB}{255,165,0}
\definecolor{frenchplum}{RGB}{129,20,83}
\definecolor{verylightgray}{rgb}{0.9, 0.9, 0.9}

\renewcommand{\arraystretch}{1.3}

\newcommand{\name}{\texttt{H-NESSi}}

\newcommand{\OO}[1]{\mathcal{O}(#1)}

\newcounter{bla}

\journal{Computer Physics Communications}

\begin{document}

\newacronym{KB}{KB}{Kadanoff-Baym}
\newacronym{GF}{GF}{Green's function}
\newacronym{GFs}{GFs}{Green's functions}
\newacronym{NEGF}{NEGF}{nonequilibrium Green's function}
\newacronym{NEGFs}{NEGFs}{nonequilibrium Green's functions}
\newacronym{dmft}{DMFT}{dynamical mean-field theory}
\newacronym{ppsc}{PPSC}{Pseudo-Particle Strong Coupling}
\newacronym{nca}{NCA}{Non-Crossing Approximation}
\newacronym{oca}{OCA}{One-Crossing Approximation}
\newacronym{KMS}{KMS}{Kubo-Martin-Schwinger}
\newacronym{BDF}{BDF}{backward differentiation formula}
\newacronym{vie}{VIE}{Volterra integral equation}
\newacronym{vide}{VIDE}{Volterra integro-differential equation}
\newacronym{vies}{VIEs}{Volterra integral equations}
\newacronym{vides}{VIDEs}{Volterra integro-differential equations}
\newacronym{mpi}{MPI}{Message passing interface}
\newacronym{BZ}{BZ}{Brillouin zone}
\newacronym{2B}{2B}{second-Born}
\newacronym{HF}{HF}{Hartree-Fock}
\newacronym{hdf5}{HDF5}{Hierarchical Data Format version 5}
\newacronym{1D}{1D}{one-dimensional}
\newacronym{sMig}{sMig}{self-consistent Migdal approximation}
\newacronym{uMig}{uMig}{unrenormalized Migdal approximation}
\newacronym{SC}{SC}{superconducting}

\renewcommand*{\acronymname}{List of Abbreviations}

\newcommand{\jk}[1]{[\textcolor{green}{JK: {#1}}]}
\newcommand{\tb}[1]{[\textcolor{red}{TB: {#1}}]}

\begin{frontmatter}



\title{\textbf{H-NESSi}: The \textbf{H}ierarchical \textbf{N}on-\textbf{E}quilibrium \textbf{S}ystems \textbf{Si}mulation package}


\author[a,b,c]{Thomas Blommel}
\author[d]{Jeremija Kovačević}
\author[e,f]{Jason Kaye}
\author[a,g]{Emanuel Gull}
\author[d]{Jakša Vučičević}
\author[h,i]{Denis Gole\v{z}}

\cortext[] {Corresponding author.\\\textit{E-mail address:} tblommel@ucsb.edu}
\address[a]{Department of Physics, University of Michigan, Ann Arbor, Michigan 48109, USA}
\address[b]{Department of Chemistry, University of California, Santa Barbara, California, USA}
\address[c]{Materials Department, University of California, Santa Barbara, California, USA}
\address[d]{Scientific Computing Laboratory, Center for the Study of Complex Systems,
Institute of Physics Belgrade, University of Belgrade, Pregrevica 118, 11080 Belgrade, Serbia}
\address[e]{Center for Computational Quantum Physics, Flatiron Institute,
162 5th Avenue, New York, New York 10010, USA}
\address[f]{Center for Computational Mathematics, Flatiron Institute, 162 5th Avenue, New York, NY 10010, USA}
\address[g]{Institute of Theoretical Physics, Faculty of Physics, University of Warsaw, Warsaw, Poland}
\address[h]{Jožef Stefan Institute, SI-1000 Ljubljana, Slovenia}
\address[i]{Faculty of Mathematics and Physics, University of Ljubljana, 1000 Ljubljana, Slovenia}

\begin{abstract}
We present \name{} (The Hierarchical Non-Equilibrium Systems Simulation package), an open-source software package for solving the Kadanoff–Baym equations (KBE) of nonequilibrium Green’s function (NEGF) theory using hierarchical low-rank compression techniques. The simulation of strongly correlated quantum systems out of equilibrium is severely limited by the cubic scaling in propagation time and quadratic memory growth associated with conventional two-time formulations. \name{} overcomes these limitations by combining high-order time-stepping schemes with hierarchical off-diagonal low-rank (HODLR) representations of the retarded and lesser Green’s functions, enabling controllable accuracy at substantially reduced computational cost and memory usage. Imaginary time quantities are efficiently represented using the discrete Lehmann representation (DLR), allowing compact and accurate treatment of thermal initial states. The implementation supports multiorbital systems, adaptive singular value truncation, and both shared-memory (OpenMP) and distributed-memory (MPI) parallelization strategies suitable for large-scale lattice calculations. The workflow closely mirrors established NEGF frameworks while introducing compression transparently into the propagation procedure. Benchmark applications to driven superconductors within dynamical mean-field theory and to the two-dimensional Hubbard model demonstrate favorable scaling compared to conventional implementations, with asymptotic time complexity significantly below the cubic scaling of uncompressed approaches. \name{} thus enables long-time and large-system nonequilibrium simulations of correlated quantum materials which were previously computationally prohibitive.\end{abstract}


\end{frontmatter}



{\bf PROGRAM SUMMARY}

\begin{small}
\noindent
{\em Manuscript Title:} {\color{black}\textbf{H-NESSi}: The \textbf{H}ierarchical \textbf{N}on-\textbf{E}quilibrium \textbf{S}ystems \textbf{Si}mulation package} 		\\
{\em Authors:} Thomas Blommel, Jeremija Kovačević, Jason Kaye, Emanuel Gull, Jakša Vučičević, Denis Gole\v{z}
                             \\
{\em Program Title:} \name{}		\\
{\em Journal Reference:}                                      \\
{\em Catalogue identifier:}                                   \\
{\em Licensing provisions:} MIT                        \\
{\em Programming language:}  C++      \\
{\em Computer:} Any architecture with suitable compilers including PCs and clusters.             \\
{\em Operating system:}    Unix, Linux, OSX       \\
{\em RAM:} Highly problem dependent  \\
{\em Classification:}                                         \\
{\em External routines/libraries:} cmake, eigen3, hdf5, fftw, libdlr, mpi, openmp   \\
{\em Nature of problem:} Solves equations of motion of time-dependent Green's functions on the Kadanoff-Baym contour within compressed representation\\
   {\em Solution method:} High-order timestepping method for integro-differential equations within the compressed representation on the Kadanoff-Baym contour.
   \\

\end{small}

\label{partI}

\section{Introduction}

The theoretical description of nonequilibrium phenomena in quantum materials and simulators requires numerical methods capable of treating strong correlations, long-time dynamics, and ultrafast driving on an equal footing. Many of the most intriguing experimental observations—such as transient superconductivity~\cite{fausti2011,mitrano2016} or long-lived metastable phases~\cite{giannetti2016ultrafast,delatorre2021,zhou2019nonequilibrium,murakami2023,basov2017towards,ligges2018ultrafast}—occur on picosecond to nanosecond timescales, motivating the development of propagation schemes that extend well beyond the femtosecond regime characteristic of intrinsic electronic processes. This large separation of timescales poses a significant computational challenge for any numerical method~\cite{marques2012fundamentals,schollwock2011density,haug2008quantum,werner2010,cohen2015,gull2011}. In this work, we focus on advancing the capabilities of nonequilibrium Green’s function (NEGF) approaches, which have emerged as a promising tool to capture rapid electronic evolution and correlations over extended timescales~\cite{kadanoff1962quantum,stefanucci2013nonequilibrium,schuler2020,aoki2014,murakami2023,sentef2013,freericks2006}. NEGFs also present a natural way around the ill-defined analytic continuation of Matsubara-domain data~\cite{Jarrell1996,sandvik1998,Yoshimi2019,fournier2020,yoon2018,Ying2022,huang2019, vucicevic2019,ZhangPronyAC}, which has proven useful in computing dynamical response functions in equilibrium~\cite{Erpenbeck_Gull_Cohen_2023,Dong_Gull_Strand_2022,kovacevic2025}. Another motivation comes from the need to study realistic multiorbital materials: despite their experimental relevance~\cite{murakami2023,perfetto2018ultrafast,molina2017ab,sangalli2019many,marini2009yambo,perfetto2018cheers}, multiorbital NEGF simulations remain limited due to the computational and memory costs associated with solving the Kadanoff–Baym equations, with long-time simulations particularly demanding~\cite{schuler2020,balzer2012nonequilibrium,balzer2011solving, Erpenbeck_Blommel_Zhang_Lin_Cohen_Gull_2024}.

A wide range of algorithmic strategies have been developed to address these challenges. The first important class consists of approximate formulations of the Kadanoff–Baym equations, such as the generalized Kadanoff–Baym ansatz (GKBA)~\cite{lipavsky1986,kalvova2019} and its G1–G2 variants~\cite{schlunzen2020,joost2020}. These approaches significantly reduce computational complexity and have enabled long-time simulations in both electron and electron–boson systems~\cite{tuovinen2020comparing,Reeves2023,karlsson2021,stefanucci2023}, although their accuracy can deteriorate in strongly correlated regimes~\cite{aoki2014,murakami2023}. Data-driven methods, including dynamic mode decomposition~\cite{yin22,yin23,reeves23}, and recurrent neural networks~\cite{bassi2024,zhu24}, yield efficient extrapolation schemes, but systematic error control remains challenging for propagation times far beyond the training interval; however, recently developed exponential approximations~\cite{Erpenbeck_Zhu_Yu_Zhang_Gerum_Goulko_Yang_Cohen_Gull_2025} show promise.

The second class consists of numerically exact approaches, which exploit specific structures of the two-time Green’s functions to reduce computational cost. Memory-cutting techniques leverage the rapid decay of the history kernel~\cite{Cohen_Rabani_2011,Cohen_Gull_Reichman_Millis_Rabani_2013,schuler2018,stahl2022,picano2021,dasari2021} and can achieve linear scaling when applicable, but they fail in regimes with long-range memory. The quantics tensor train (QTT) scheme~\cite{shinaoka2023,murray2024,sroda2024} provides an alternative global-in-time perspective, though convergence issues at long times require patching~\cite{inayoshi2025} or predictor–corrector methods~\cite{sroda2025}. Hierarchical matrix strategies, including hierarchical off-diagonal low-rank (HODLR) compression~\cite{kaye2021,blommel25}, offer another robust approach, and the possibility of combining compression with timestepping. These methods provide controllable accuracy and a systematic, model-agnostic route to reducing computational cost and memory usage by exploiting low-rank structures in two-time Green’s functions. There is also the possibility of combining them with global iteration~\cite{lamic2024,gasperlin25}, enabling fast hierarchical matrix algebra~\cite{ballani16}.

In this paper, we present an open-source and high-order accurate implementation of the HODLR scheme presented in Ref.~\cite{kaye2021}, generalized to multiorbital systems. It features adaptive rank selection and user-controlled accuracy without introducing additional approximations beyond the hierarchical compression itself. The library offers a flexible workflow that can be readily integrated with applications such as dynamical mean-field theory (DMFT)~\cite{georges1996dynamical,aoki2014,murakami2023} and impurity solvers. The code was designed with a user experience similar to the NESSi package~\cite{schuler2020} while enabling multiorbital simulations at significantly longer propagation times. We illustrate the capabilities of the library with benchmark calculations on driven superconductors within DMFT following Ref.~\cite{blommel2024}, as well as with the second Born approximation for the two-dimensional Hubbard model. The latter example follows Ref.~\cite{kovacevic2025} and combines the solution of the Kadanoff–Baym equations with efficient parallelization schemes, enabling simulations of large lattice systems that preserve translation invariance.

\section{Overview of multiorbital HODLR timestepping\label{sec:overview}}
\begin{figure}
    \centering
    \centerline{\includegraphics[width=1.5\linewidth]{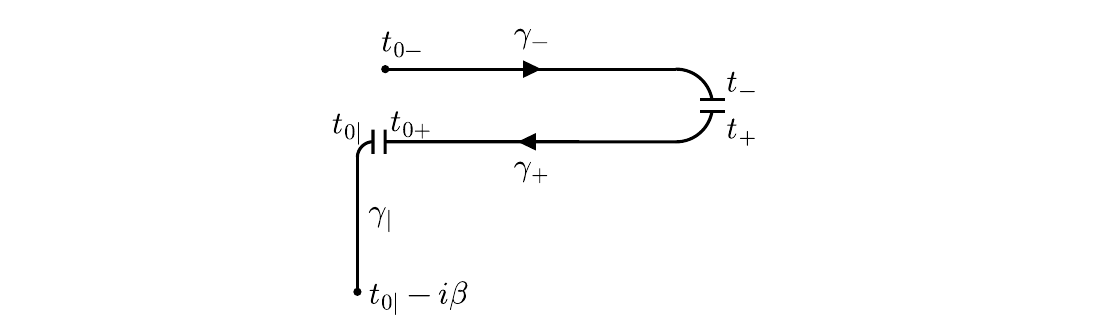}}
    \caption{The Konstantinov-Perel' contour.}
    \label{fig:Contour}
\end{figure}

The non-equilibrium single-particle Green's function is defined as 
\begin{equation}
    G_{ij}(z,z') = \frac{1}{i}\frac{\mathrm{Tr}\left[\mathcal{T}\left\{e^{-i\int_\gamma d\bar{z}\hat{H}(\bar{z})}\hat{d}_i(z)\hat{d}^\dagger_j(z')\right\}\right]}{\mathrm{Tr}\left[\mathcal{T}\left\{e^{-i\int_\gamma d\bar{z}\hat{H}(\bar{z})}\right\}\right]}, \label{eq:Gdef}
\end{equation}
where $\gamma$ is the Konstantinov-Perel' contour~\cite{konstantinov1961diagram} shown in Fig.~\ref{fig:Contour}, $\mathcal{T}$ is the contour ordering operator, and $\hat{d}^\dagger_j(z')$ creates a particle in orbital $j$ at contour time $z'$. In what follows, we suppress the orbital indices; all propagators should be understood as matrices in orbital space. It is useful to define Keldysh components---functions of real-valued arguments in which the contour locations are made explicit---as shown in Table~\ref{tab:BlockStructureG}.
\begin{table}[h!]
  \centering
  {\renewcommand{\arraystretch}{1.25} 
  \begin{tabular}{|c|c|}
    \hline  Component & Name \\
    \hline $G^{<}(t,t') = G(t_-,t_+)$ & lesser \\
    \hline $G^{>}(t,t') = G(t_+,t_-)$ & greater \\
    \hline $G^{\rceil}(t,\tau) = G(t_\pm,t_0-i\tau)$ & left-mixed \\
    \hline $G^{\lceil}(\tau,t) = G(t_0-i\tau,t_\pm)$ & right-mixed \\
    \hline $G^M(\tau) = -iG(-i\tau,t_0)$ & Matsubara \\
    \hline
  \end{tabular}
  }
\caption{Keldysh components of the Green's function and their locations on the contour.  $t_\pm$ explicitly places the argument on the $\gamma_\pm$ horizontal leg of the contour.}
\label{tab:BlockStructureG}
\end{table}
It is also convenient to define the retarded and advanced components
\begin{align}
    G^R(t,t') &= \theta(t-t')[G^>(t,t')-G^<(t,t')]\label{eq:GRdef}\\
    G^A(t,t') &= \theta(t'-t)[G^<(t,t')-G^>(t,t')].
\end{align}
These Keldysh components are not all independent; a minimal set of four suffices to capture all information in the single-particle Green's function.  There are multiple possible choices; following Ref.~\cite{schuler2020}, we choose the components $\{M,\rceil,R,<\}$ and refer to the left-mixed component simply as the mixed component.  From this set, we can obtain the advanced and right-mixed components from
\begin{align*}
    G^A(t,t') &= [G^R(t',t)]^\dagger\\
    G^\lceil(\tau,t)&=-\xi[G^\rceil(t,\beta-\tau)]^\dagger
\end{align*}
where $\xi=\pm$ for bosons (upper sign) and fermions (lower sign), and $\beta$ is the inverse temperature.
We only need the lesser and greater components on one side of the $(t,t')$ diagonal, due to their anti-Hermitian symmetry
\begin{align}
    G^\gtrless(t,t') &= -[G^\gtrless(t',t)]^\dagger.\label{eq:GLantiHerm}
\end{align}
Finally, all imaginary time functions need only be known on $\tau\in[0,\beta)$, due to (anti-)periodicity
\begin{align*}
    G^M(\tau) &= \xi G^M(\tau+\beta)\\
    G^\rceil(t,\tau) &= \xi G^\rceil(t,\tau+\beta).
\end{align*}

In this work, we discretize the time axis into equidistant points with spacing $h$.  The lesser and retarded components are functions of two times and can therefore be viewed as matrices with entries $G^R_{t_1t_2} = G^R(t_1h,t_2h)$ and $G^<_{t_1t_2} = G^<(t_2h,t_1h)$ (here suppressing the orbital indices).  We often drop the factor of $h$ and write $G(t_1,t_2)$ for the corresponding matrix element, with integers $t_1,t_2$.  The definition of the retarded component in Eq.~\ref{eq:GRdef} contains a Heaviside function that restricts this matrix to being lower triangular.  The lesser component can similarly be stored as a lower triangular matrix, as the anti-Hermitian symmetry \eqref{eq:GLantiHerm} allows us to recover information above the diagonal.  
For further details on the numerical discretization, anti-Hermitian symmetries, and related aspects, we refer the reader to Ref.~\cite{schuler2020}. These triangular representations apply equally to Green's functions, self-energies, and hybridization functions in the context of DMFT.

\subsection{Kadanoff-Baym equations}
The KBE are a coupled set of integro-differential equations for the Keldysh components of the Green's function, here written in matrix form:
\begin{align}
    &-\partial_\tau G^M(\tau) = \delta(\tau)\mathbf{1}+\epsilon(0_|)\cdot G^M(\tau)\label{eq:KBEM}\\&+\int_0^\beta d\bar{\tau}\Sigma^M(\tau-\bar{\tau}) \cdot G^M(\bar{\tau})\nonumber\\
    &i\partial_{t'}G^R(t,t-t') = G^R(t,t-t')\cdot \epsilon(t-t') \label{eq:KBER}\\&+ \int_0^{t'}d\bar{t} G^R(t,t-\bar{t})\cdot \Sigma^R(t-\bar{t},t-t')\nonumber\\
    &i \partial_t G^{\rceil}(t, \tau)=\epsilon(t)\cdot G^{\rceil}(t, \tau)+\int_0^t d \bar{t} \Sigma^R(t, \bar{t}) \cdot G^{\rceil}(\bar{t}, \tau)\label{eq:KBET}\\&+\int_0^\beta d \bar{\tau} \Sigma^{\rceil}(t, \bar{\tau}) \cdot G^M(\bar{\tau}-\tau)\nonumber\\
    &i\partial_t G^<(t,t') =\epsilon(t)\cdot G^<(t,t') + \int_0^td\bar{t}\Sigma^R(t,\bar{t}) \cdot G^<(\bar{t},t')\label{eq:KBEL}\\&+\int_0^{t'}d\bar{t}\Sigma^<(t,\bar{t}) \cdot G^A(\bar{t},t') -i\int_0^\beta d\bar{\tau}\Sigma^\rceil(t,\bar{\tau}) \cdot G^\lceil(\bar{\tau},t').\nonumber
\end{align}
Here $\cdot$ marks matrix multiplication in the orbital space and $\epsilon(0_|)$ denotes the equilibrium value of $\epsilon(t)$ on the thermal branch.  The function $\epsilon(t)=h_0(t)+\Sigma^{MF}(t)-\mu$ is the quadratic mean-field Hamiltonian, which contains the quadratic terms of the Hamiltonian, the mean-field self-energy, and the chemical potential, respectively. 

The Keldysh components of the correlated (beyond mean-field) self-energy $\Sigma$ obey the same symmetries as the Green's functions.  Since the self-energy is typically a nonlinear functional of the Green's function, the KBE Eqs.~\ref{eq:KBEM}-\ref{eq:KBEL} must be solved self-consistently.  We use an iterative scheme in which, at iteration $i$, the self-energy is evaluated from the current Green's function, $\Sigma^{(i)}[G^{(i)}]$, and the KBE are then solved to obtain the updated $G^{(i+1)}[\Sigma^{(i)}]$.  Following Ref.~\cite{schuler2020}, the nonlinear equation is converged self-consistently at each timestep before proceeding to the next, except during bootstrapping, where global iterations are used (see Secs.~\ref{ssec:dyson_class} and \ref{ssec:example_boot}).  In practice, we find this procedure to be very robust, in contrast to global iterative schemes which can require considerable effort to stabilize~\cite{Dong_Krivenko_Kleinhenz_Antipov_Cohen_Gull_2017,sroda2024,inayoshi2025,sroda2025,gasperlin25,picano2025}.

\subsection{Hierarchical decomposition}
\begin{figure}
    \centering
    \centerline{\includegraphics[width=1.8\linewidth]{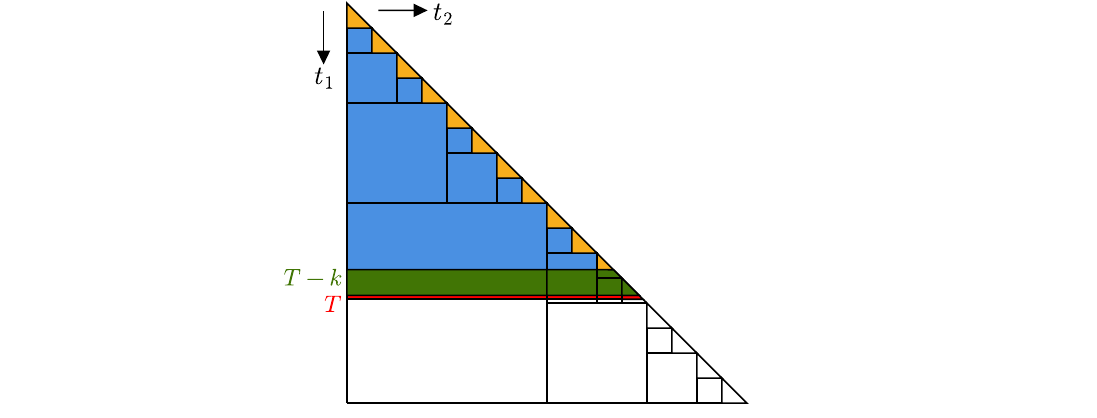}}
    \caption{Snapshot of HODLR representation of $C^R(t_1,t_2)$ and $C^<(t_2,t_1)$ for $C\in\{G,\Sigma\}$ at timestep $T$.  The blue areas are stored in the truncated SVD representation; all other colored regions are stored directly.  The horizontal red slice is the timestep currently being solved for.  The green region consists of the previous $k$ timesteps, which are stored directly and will eventually be used to update the blue and yellow regions.  White regions are yet to be solved for.}
    \label{fig:hodlr}
\end{figure}

The favorable scaling of \name{} is a consequence of the use of compressed representations of the two-time Green's functions and self-energies~\cite{kaye2021}, achieved through a hierarchical partitioning of the lower-triangular retarded and lesser components, as shown in Fig.~\ref{fig:hodlr}.  The user supplies the number of timesteps $N_t$ and hierarchical levels $N_l$, which together determine the partitioning, and these parameters remain fixed throughout the calculation.  In the example shown, $N_l=4$, as evidenced by the four distinct block sizes.  In general, a level $l$ consists of $2^{l-1}$ blocks, each of approximate size $\frac{N_t}{2^l}$.  If $\frac{N_t}{2^l}$ is not an integer, some blocks will be rectangular rather than square; however, this does not affect the efficiency of the algorithm or the user interface.  

For efficient storage, each block $B$ is stored as a truncated SVD (TSVD) $B\approx USV^\dagger$, where the diagonal matrix $S$ contains singular values and $U$, $V$ are the corresponding singular vectors. Singular values below a user-provided threshold $\varepsilon_{SVD}$, along with their associated vectors, are discarded, yielding accuracy $|B-USV|<\varepsilon_{SVD}$.  The number of retained singular values, $N_S^\varepsilon$, is the $\varepsilon$-rank of the block, which is typically much smaller than the block size. These ranks often grow sublinearly with block size, improving compression efficiency at long integration times.

Integrating the KBE amounts to incrementally filling in rows of the lower triangular matrices of $G^<$ and $G^R$.  This is shown in Figure \ref{fig:hodlr}, where the red row at timestep $T$ is the current row being solved for, and all the rows above have already been computed.  In the illustrated example, two blue blocks are partially built, as only some of their rows have been calculated.  As described in Ref.~\cite{kaye2021}, these blocks are still stored as a truncated SVD.  As new rows are obtained via integration of the KBE, the truncated SVD is incrementally updated using an efficient rank-1 update algorithm.

\subsection{Multiorbital compression}
The Green's function and self-energy each carry two indices enumerating degrees of freedom such as spin, orbital, and lattice site (the latter is needed for inhomogeneous lattice models; otherwise a single momentum index suffices).  We refer to these as orbital indices and denote their number by $N_o$.  Multiorbital compression is accomplished by separately storing $N_o^2$ compressed representations of the two-time functions.  The complexity of managing these separate compressed functions is hidden from the user, as can be seen in Tables \ref{tab:getset}-\ref{tab:getmat_offnode}, which show $N_o\times N_o$ matrices passed between the user and the library. Offline tests on simple multiorbital systems indicate that more sophisticated compression techniques combining orbital and time indices do not yield significant efficiency improvements over the current approach and can reduce compression efficiency while increasing the complexity of evaluating history integrals.

\subsection{Numerical integration}
At timestep $T$, we must solve for the retarded, mixed, and lesser components.  This requires us to approximate the derivatives and history integrals evaluated on an equidistant grid with spacing $h$.  These approximations are implemented in the \texttt{Integration} class, which supports discretization error scaling as $h^{k+1}$, where $1\leq k\leq5$ is a user-specified parameter.  In \ref{app:integration}, we provide a brief discussion of the approximation schemes we use for evaluating derivatives and integrals.  For an in-depth discussion on how these high-order integrators are implemented, we refer the reader to Refs.~\cite{schuler2020,blommelthesis}.

\subsection{Discrete Lehmann representation}
\label{ssec:DLR}
\name{} uses the discrete Lehmann representation (DLR)~\cite{kaye22}, via the \texttt{libdlr} library~\cite{KAYE_libdlr}, to discretize imaginary time functions on $[0,\beta]$.  The \texttt{dlr\_info} class provides a simple interface, managing the internal data and accessor functions. We give an overview of the method and its user-facing parameters.

The \texttt{libdlr} library requires two parameters, $\Lambda$ and $\epsilon_{\text{dlr}}$, which control the accuracy of the representation.  $\Lambda=\beta \omega_\text{max}$ sets the maximum spectral support, scaled by the inverse temperature; it can typically be estimated from the energy scales of the system under study.  The second parameter, $\epsilon_{\text{dlr}}$, sets the numerical precision of the representation.  Given these parameters, the DLR is constructed by selecting a sparse grid $\tau_i\in(0,\beta)$ on which the imaginary time axis is sampled. The number of sampling points $r$ has been shown to scale as $r=\OO{\log{(\Lambda)}\log{(1/\epsilon_\text{dlr})}}$~\cite{kaye22}. From these nodes, a sum-of-exponentials representation of the function---the DLR expansion---can be constructed.

\name{} stores all imaginary time functions exclusively on this sparse DLR grid, so the user is responsible for evaluating the self-energy at these nodes.  Helper functions in the \texttt{dlr\_info} and \texttt{herm\_matrix\_hodlr} classes enable evaluation at arbitrary $\tau\in[0,\beta]$; see Tables~\ref{tab:getmat_onnode}~and~\ref{tab:getmat_offnode} for examples.  Example self-energy evaluation functions for the Matsubara and mixed components are provided in \ref{app:selfen}.

\section{Features and layout of the library\label{sec:basic}}
Here we list the main capabilities of \name:

\begin{itemize}
    \item Store and incrementally build two-time finite-temperature Green's
    functions and self-energies in compressed representation using
    the \texttt{herm\_matrix\_hodlr} class
    \item Numerically integrate the KBE, leveraging the compressed representations of $G$ and $\Sigma$, using the \texttt{dyson} class, with first- to sixth-order integrators implemented in the \texttt{integration} class
    \item Manipulate imaginary-time Green's functions in the DLR format using the \texttt{DLR} class
    \item Store one-time contour functions in the \texttt{function} class
\end{itemize}

\name{} is implemented in C++ and includes Python scripts for reading, plotting, and profiling.  The documentation provides build instructions, API information, and example programs.  The library is available as a public Git repository at \url{https://github.com/KBE-hodlr/H-NESSi}.  The folder \texttt{\name/test} contains unit tests to verify installation and output of example programs.  \name{} uses the \texttt{Eigen} linear algebra package to perform linear algebra routines and interface with the data structures.  All data is stored using row-major ordering.  

The \name{} interface is designed to be nearly identical to the NESSi interface~\cite{schuler2020}. The main difference is the underlying compressed representation of the two-time quantities. As long as the user only modifies or retrieves values at the current timestep, a \name{} program is essentially the same as a NESSi one.  Retrieval of compressed data is possible through the same interface, but the reconstruction is expensive, and can be avoided in almost all cases.  The second difference is the use of the DLR grid instead of equidistant imaginary-time points, which requires minimal code modifications via the functions in Tables \ref{tab:getmat_onnode} and \ref{tab:getmat_offnode}.  

The final difference is the introduction of four additional parameters, which we now list along with guidelines for their selection.  $N_l$ is the number of levels in the hierarchical decomposition of the Green's functions and self-energies. In Fig.~\ref{fig:hodlr} we have $N_l=4$, as the decomposition contains four different block sizes.  We suggest choosing this value so that $N_t/2^{N_l} \approx 10$.  Next is the SVD truncation parameter, $\varepsilon_{SVD}$ which represents the desired precision of the solution.  In our experience, 
$10^{-5}<\epsilon_{SVD}$ can lead to amplification of errors due to inaccurate evaluation of history integrals.  For small values of $\epsilon_{SVD}$, care must be taken to ensure that noise arising from unconverged self-consistency loops or discretization error from the finite integration order does not enter the compressed representations.
Lastly, the two DLR parameters, $\epsilon_{\rm dlr}$ and $\Lambda$, are discussed in Sec.~\ref{ssec:DLR}.

\subsection{The {\tt herm\_matrix\_hodlr} class}
In this section, we provide an overview of functionality contained within the {\tt herm\_matrix\_hodlr} class, including accessing and updating the data it contains, as well as updating the compressed representations at each timestep.
\texttt{herm\_matrix\_hodlr} is the class responsible for storing two-time Green's functions and self-energies used in the integration of the KBE.  Each object stores four Keldysh components: Matsubara ($M$), retarded ($R$), lesser ($<$), and mixed ($\rceil$). The Matsubara component is stored on the DLR grid. The mixed component has two arguments: the imaginary axis is discretized on the DLR grid, and the real axis on an equispaced grid.  Neither of these two components are stored in an SVD-compressed format. 
The retarded and lesser components are stored in the HODLR-compressed representation; see Fig.~\ref{fig:hodlr}.
A core responsibility of the \texttt{herm\_matrix\_hodlr} class is managing the HODLR representation and updating the TSVD of each block as new rows are computed.  This is done by calling \texttt{update\_blocks} once per timestep.

Because of the high-order integration routines, the compressed representation lags the timestepping by $k+1$ steps, where $k$ is the integration order.  At timestep $T$, the data $G^R(T-k\leq t\leq T,0\leq t'\leq t)$ and $G^<(0\leq t\leq t', T-k\leq t'\leq T)$---corresponding to the red and green regions in Figure \ref{fig:hodlr}---has been computed but not yet compressed.  These regions are stored directly, allowing efficient access and modification via the functions in Table~\ref{tab:getset}, which provide \texttt{Eigen} maps or copy data to/from \texttt{Eigen} matrices.  These functions will fail if used to access data outside the red and green regions.  

The blue and yellow regions of Figure \ref{fig:hodlr} correspond to data stored in TSVDs and directly stored triangles, respectively; these are read-only and must be accessed via the functions in Table \ref{tab:getcompress}, valid for any $(t,t')$ up to the current timestep $T$.  For points in the yellow, green, and red regions, these functions simply copy directly stored data.  Accessing data in the TSVD (blue) region requires reconstruction of the function values via the TSVD, which requires $\OO{N_S^\varepsilon}$ operations.  If used often, the cost of this reconstruction procedure can become significant compared to cases in which data is only accessed while in the directly stored regions.
Solving the KBE only requires the user to modify the self-energy at the current timestep (red region), meaning that for almost all cases, reconstruction of Green's function and self-energy values from the TSVD can be avoided.

\begin{table}[h!]
    \centering
    \begin{tabular}{|c|c|}\hline
        \makecell{\texttt{get\_les\_curr}\\
        \texttt{set\_les\_curr}}& \makecell{Copy $N_o\times N_o$ lesser data $C^<_{ij}(t,t')$ \\ at given $(t,t')$ to/from \\ provided \texttt{Eigen} matrix.} \\\hline
        \texttt{map\_les\_curr} & \makecell{Return $N_o\times N_o$  \texttt{Eigen} matrix map\\ of $C^<_{ij}(t,t')$ at given $(t,t')$} \\\hline
        \makecell{\texttt{get\_ret\_curr}\\
        \texttt{set\_ret\_curr}} & \makecell{Copy $N_o\times N_o$ retarded data $C^R_{ij}(t,t')$ \\ at given $(t,t')$ to/from \\ provided \texttt{Eigen} matrix.} \\\hline
        \texttt{map\_ret\_curr} & \makecell{Return $N_o\times N_o$  \texttt{Eigen} matrix map\\ of $C^R_{ij}(t,t')$ at given $(t,t')$} \\\hline
    \end{tabular}
    \caption{get, set, and map functions for the retarded and lesser Keldysh components in the red and green regions of Fig.~\ref{fig:hodlr}.  These functions will fail if used outside of the red or green regions of Fig.~\ref{fig:hodlr}.}
    \label{tab:getset}
\end{table}
\begin{table}[h!]
    \centering
    \begin{tabular}{|c|c|}\hline
        \texttt{get\_les} & \makecell{Evaluate compressed representation of $C^<_{ij}(t,t')$ \\ at given $(t,t')$ into provided \texttt{Eigen} matrix.} \\\hline
        \texttt{get\_ret} & \makecell{Evaluate compressed representation of $C^R_{ij}(t,t')$ \\ at given $(t,t')$ into provided \texttt{Eigen} matrix.} \\\hline
    \end{tabular}
    \caption{General get functions for retarded and lesser components that are valid for any $(t,t')$ point up to the current timestep.  For a point in the blue region of Fig.~\ref{fig:hodlr}, this involves reconstructing the value from the TSVD representation.  Points in the yellow, green, and red regions simply involve copying data. }
    \label{tab:getcompress}
\end{table}

As discussed in Sec.~\ref{ssec:DLR}, imaginary-time functions are stored on the sparse DLR grid $\tau_i\in(0,\beta)$.  The functions in Table \ref{tab:getmat_onnode} allow retrieval and manipulation of this data.
It is often necessary, however, to evaluate the Matsubara and mixed components at points off the DLR grid---for example, to compute the equilibrium density matrix $-G^M(\beta^-)$ (the DLR grid does not include endpoints).  The functions in Table~\ref{tab:getmat_offnode} evaluate imaginary-time functions at arbitrary $0\leq\tau\leq\beta$ using the DLR expansion.  One can also obtain imaginary-time quantities on the ``reversed'' DLR grid $\tau_i\rightarrow\beta-\tau_i$, which is needed for self-energy diagrams involving quantities such as $G^M(-\tau)=\xi G^M(\beta-\tau)$.
\begin{table}[h!]
    \centering
    \begin{tabular}{|c|c|}\hline
        \makecell{\texttt{get\_mat}\\\texttt{set\_mat}} & \makecell{Copy $N_o\times N_o$ Matsubara data $C^M_{ij}(\tau)$ \\ at given DLR node $\tau_i$ to/from \\ provided \texttt{Eigen} matrix.} \\\hline
        \texttt{map\_mat} & \makecell{Return $N_o\times N_o$ \texttt{Eigen} matrix map \\of $C^M_{ij}(\tau)$ at given DLR node $\tau_i$ } \\\hline
        \makecell{\texttt{get\_tv} \\\texttt{set\_tv}}& \makecell{Copy $N_o\times N_o$ mixed data $C^\rceil_{ij}(t,\tau)$ \\ at given DLR node $\tau_i$ and timestep $t$\\ to/from provided \texttt{Eigen} matrix.} \\\hline
        \texttt{map\_tv} & \makecell{Return $N_o\times N_o$ \texttt{Eigen} matrix map \\of $C^\rceil_{ij}(t,\tau)$ at given \\DLR node $\tau_i$ and timestep $t$} \\\hline
    \end{tabular}
    \caption{Functions to access imaginary time functions at DLR nodes.}
    \label{tab:getmat_onnode}
\end{table}

\begin{table}[h!]
    \centering
    \begin{tabular}{|c|c|}\hline
        \texttt{get\_mat\_tau} & \makecell{Evaluate DLR expansion of $C^M_{ij}(\tau)$\\ for arbitrary $0\leq\tau\leq\beta$ \\ into provided \texttt{Eigen} matrix.} \\\hline
        \texttt{get\_tv\_tau} & \makecell{Evaluate DLR expansion of $C^\rceil_{ij}(t,\tau)$\\ for arbitrary $0\leq\tau\leq\beta$ and given  \\ timestep into provided \texttt{Eigen} matrix.} \\\hline
        \texttt{get\_mat\_reversed} & \makecell{Evaluate $C^M_{ij}(\beta-\tau_l)$ into provided \\$N_\tau\times N_o^2$ \texttt{Eigen} matrix, \\where $\tau_l$ are DLR nodes} \\\hline
        \texttt{get\_tv\_reversed} & \makecell{Evaluate $C^\rceil_{ij}(t,\beta-\tau_l)$ into provided \\$N_\tau\times N_o^2$ \texttt{Eigen} matrix, \\where $\tau_l$ are DLR nodes} \\\hline
    \end{tabular}
    \caption{Functions to evaluate DLR representations at points not on DLR nodes.  }
    \label{tab:getmat_offnode}
\end{table}

\subsection{The {\tt dyson} class}
\label{ssec:dyson_class}
The \texttt{dyson} class solves the KBE in three steps: (1) solving the Matsubara problem, (2) bootstrapping, and (3) timestepping.
In this section we will discuss the functions provided that solve the KBE for the respective steps.  In all three cases, these functions will return the norm $|G^{(i)}-G^{(i-1)}|$ between successive self-consistent iterations, which can be used to terminate the loop once the desired accuracy is reached.

First, the Matsubara problem is solved to obtain the initial thermal state.  This step is optional: systems may alternatively be initialized via adiabatic switching on the two-legged contour or with an uncorrelated density matrix, in which case the user begins directly at bootstrapping.
\begin{table}[h!]
\centering
    \begin{tabular}{|c|c|} \hline
        function & description  \\ \hline
        \texttt{dyson\_mat} & Perform a Matsubara iteration.   \\ \hline
    \end{tabular}
    \caption{Matsubara functions in the \texttt{dyson} class}
    \label{tab:Matdyson}
\end{table}

The bootstrap process initializes the $k^\text{th}$-order accurate multistep timestepper by handling the first $k$ timesteps.  The first $k$ timesteps are solved simultaneously, meaning that at each iteration the user must update $h^{MF}$ and $\Sigma$ for all $0\leq t\leq k$ before calling one of the bootstrap functions listed in Table~\ref{tab:Bootdyson}.  To initialize the self-consistency loop, we must provide an initial guess for $G$.  This can be done using the \texttt{green\_from\_H} function, which evaluates the mean-field real-time Green's function for the first $k$ timesteps.

If the system is driven out of equilibrium in the first $k$ timesteps, for example, by an interaction quench or an electric field pulse, the user should use the non-time-translation invariant (ntti) solver \texttt{dyson\_start\_ntti}, which makes no assumptions regarding the dynamics of the system, and directly solves the KBE.  This function implements the same bootstrap procedure as laid out in Ref.~\cite{schuler2020}.  If the system remains in the translationally invariant state relative to the thermal preparation for at least the first $k$ timesteps, we provide the function \texttt{dyson\_start}, which only solves the mixed component equation and uses time-translation invariance to obtain $G^R$ and $G^<$.  Lastly, the function \texttt{dyson\_start\_2leg} is for cases when the system is not prepared in an initial thermal state and therefore has no imaginary-time leg.  In such cases, only the retarded and lesser components are solved.
\begin{table}[h!]
\centering
    \begin{tabular}{|c|c|} \hline
        function & description  \\ \hline
        \texttt{dyson\_start} & \makecell{Perform a Bootstrap iteration. \\ $\Sigma$ and $h^{MF}$ must be TTI for first \\ $k$ timesteps.}  \\ \hline
        \texttt{dyson\_start\_ntti} & \makecell{Perform a Bootstrap iteration. \\ No requirements on $\Sigma$ or $h^{MF}$.} \\ \hline
        \texttt{dyson\_start\_2leg} & \makecell{Perform a Bootstrap iteration. \\ Solve without $\rceil$ or $M$ component.} \\ \hline
    \end{tabular}
    \caption{Bootstrap functions in the \texttt{dyson} class.}
    \label{tab:Bootdyson}
\end{table}

After the bootstrap is complete, the system is integrated using the timestepping functions. Before each timestep, an initial guess for the self-consistency loop may be obtained by extrapolating from the previous $k$ timesteps using the \texttt{extrapolate} function.  After this, the timestep may be solved self-consistently using one of the two timestep functions shown in Table~\ref{tab:Timedyson}.  
\begin{table}[h!]
\centering
    \begin{tabular}{|c|c|} \hline
        function & description  \\ \hline
        \texttt{dyson\_timestep} & \makecell{Perform a timestep iteration. \\ No requirements on $\Sigma$ or $h^{MF}$.}  \\ \hline
        \texttt{dyson\_timestep\_2leg} & \makecell{Perform a timestep iteration. \\ Solve without $\rceil$ or $M$ component.} \\ \hline
    \end{tabular}
    \caption{Timestepping functions in the \texttt{dyson} class.}
    \label{tab:Timedyson}
\end{table}

\begin{figure*}
    \centering
    \includegraphics[width=\linewidth]{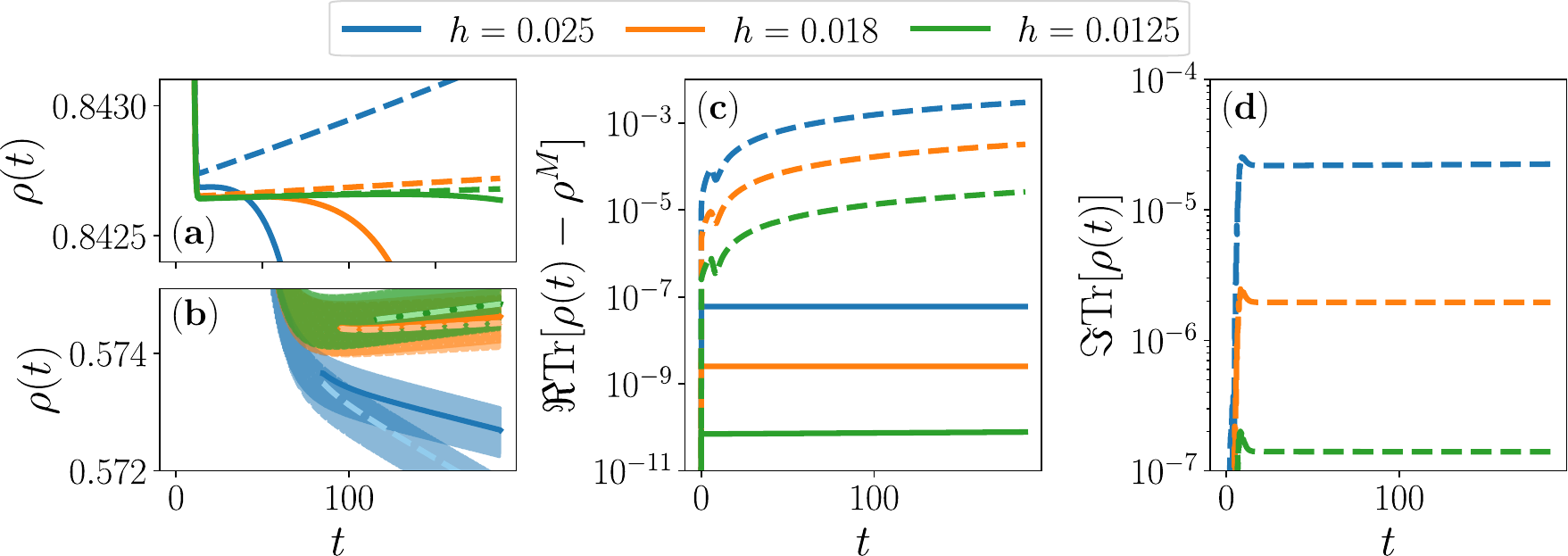}
    \caption{Comparison of the dynamics for a diagonal element in the density matrix for \texttt{RHO\_DIAGONAL} (solid lines) and \texttt{RHO\_HORIZONTAL} (dashed lines) after (a) a short pulse excitation and (b) continuous periodic driving. The solid dark lines and light dashed lines overlaid in (b) are rolling averages meant to guide the eye and demonstrate convergence, as the underlying data is highly oscillatory. Real (c) and imaginary (d) part of the density matrix trace for continuous driving.  All calculations use $\epsilon_{SVD}=10^{-10}$.}
    \label{fig:d_vs_h}
\end{figure*}

The density matrix, $\rho(t) = \xi iG^<(t,t)$, describes the particle distribution, and enters the KBE via the mean-field Hamiltonian, making its accurate propagation essential. When constructing the dyson class, one of two arguments must be provided: either \texttt{RHO\_HORIZONTAL} or \texttt{RHO\_DIAGONAL}. The former provides no special treatment of the density matrix and simply integrates the lesser Green's function as laid out in Ref.~\cite{schuler2020}. The latter treats the density matrix in a special manner by integrating the equation
$$\partial_t\rho(t)|_{t=T} = \xi i\partial_tG^<(t,T) + \xi i\partial_tG^<(T,t)$$ along the $t=t^{\prime}$ diagonal, which incurs no extra computational expense. While both schemes solve the KBE, the diagonal integration provided by \texttt{RHO\_DIAGONAL} often results in more strictly preserved conservation laws.

To compare the two schemes, we plot population dynamics for an initially occupied site in a simple two-band, two-site Hubbard model in Fig.~\ref{fig:d_vs_h}. The state is initially fully occupied before a driving field is turned on at $t=10$ to excite a portion of the population into the conduction band.  We consider two variations of this setup: a short pulse excitation where the field is quickly turned off, Fig.~\ref{fig:d_vs_h}a, and continuous periodic driving leading to sustained population oscillations, Fig.~\ref{fig:d_vs_h}b.  In both instances, the two methods converge to the same physical result as the timestep size $h$ is reduced.  In the continuous driving case, this convergence is demonstrated by the rolling averages for $h=0.0125$, for which the diagonal method (solid dark line) and the horizontal method (dashed light line) perfectly overlap.  

Advantages of the diagonal integration are shown in (c), where we demonstrate that the diagonal integration very accurately preserves the total number of particles in the system, whereas the horizontal integration is worse by several orders of magnitude and has a continually compounding loss of particle number. Furthermore, (d) illustrates that the horizontal integration method causes a non-zero imaginary component in the density matrix, which in turn leads to a small non-Hermitian part of the mean-field Hamiltonian. While one could manually force the mean-field Hamiltonian to be Hermitian at each timestep, this ad-hoc intervention only corrects the time-evolution operator; it does not fix the underlying non-Hermiticity of the density matrix itself, meaning physical observables will still be corrupted. The diagonal method naturally avoids this flaw entirely.

A notable caveat to and failure of the diagonal method occurs in steady-state regimes; in situations where the system remains in a steady-state for long periods of time, as in panel (a), the diagonal integration may fail and rapidly diverge. Therefore, in simulation regimes where the system is expected to remain in a steady state for long times, the diagonal method should be avoided, and \texttt{RHO\_HORIZONTAL}, with a manual Hermitian correction to the mean-field Hamiltonian, should be used instead.

\section{Example program: driven superconductor} \label{sec:example_programs}
Here we provide an example implementation that showcases how to use \name{} to integrate the KBE.  We solve a driven superconducting system modeled using the attractive Hubbard model, which is solved self-consistently using the DMFT approximation and a second-Born self-energy.  We provide a brief overview of the problem setup; a more in-depth discussion is available in Ref.~\cite{blommel25}.

The attractive Hubbard Hamiltonian is given by 
\begin{equation}\label{Eq:Model}
     H = -t_0 \sum_{\langle j,k\rangle\sigma}e^{iqA(t)} d^\dagger_{j\sigma}d_{k\sigma}-U\sum_j (n_{j\uparrow}-1/2)(n_{j\downarrow}-1/2),
\end{equation}
where $d_{i\sigma}$ is the annihilation operator at site $i$ and spin $\sigma$,
$n_{i\sigma}$ is the spin-dependent density operator and $q$ is the charge. To describe the dynamics  in the superconducting state, it is convenient to rewrite the problem using Nambu spinors 
\begin{gather}
    \psi_i = \begin{pmatrix}
        d_{i\uparrow}\\
        d^\dagger_{i\downarrow}
    \end{pmatrix}.
\end{gather}
The Hamiltonian now takes the form 
\begin{align}
H(t) &= -t_0 \sum_{\langle j,k\rangle\alpha}\alpha e^{iqA(t)\alpha}\psi^\dagger_{j\alpha}\psi_{k\alpha}\\&+U(t)\sum_i \psi^{\dagger}_{i+}\psi_{i+}\psi^{\dagger}_{i-}\psi_{i-}-\frac{U(t)}{2}\sum_{i\alpha}\psi^\dagger_{i\alpha}\psi_{i\alpha},\nonumber
\end{align}
where $\alpha\in\{-1,1\}$ runs through the first and second components of the Nambu spinor.
We solve the KBE for the normal and anomalous Green's functions 
\begin{equation}
    G(t,t') = \langle\psi(t)\psi^\dagger(t')\rangle = \begin{pmatrix}
        \langle d_{\uparrow}d^\dagger_{\uparrow}\rangle(t,t') && \langle d_{\uparrow}d_{\downarrow}\rangle(t,t') \\
        \langle d^\dagger_{\downarrow}d^\dagger_{\uparrow}\rangle(t,t') && \langle d^\dagger_{\downarrow}d_{\downarrow}\rangle(t,t')
    \end{pmatrix}.
\end{equation}
The system is then mapped to a single site using DMFT on a Bethe lattice, which introduces the hybridization function for the driven situation~\cite{li2019long},
\begin{equation}
\begin{split}
    \Delta(t,t') = \Delta_R(t,t') + \Delta_L(t,t')\\
    \Delta_R(t,t') = \frac{1}{2}\Bar{t}_0(t)\sigma_z G(t,t') \sigma_z \Bar{t}_0^*(t')\\
    \Delta_L(t,t') = \frac{1}{2}\Bar{t}_0^*(t)\sigma_z G(t,t') \sigma_z \Bar{t}_0(t'),
    \label{eq:Hybrid}
\end{split}
\end{equation}
with hopping matrix elements given by
\begin{align}
    \Bar{t}_0(t) &= \begin{pmatrix}
        e^{iA(t)} && 0 \\
        0 && e^{-iA(t)}
    \end{pmatrix},\\
    A(t) &= -\int_0^t E(s) \, ds,
\end{align}
due to the Peierls substitution. Here $E$ is the driving electric field $
    E(t) = E_0 \, \text{exp}\left(-\frac{(t-t_c)^2}{\sigma^2}\right)\text{sin}(\omega (t-t_c)),
$
with pulse center $t_c$, pulse width $\sigma$, driving frequency $\omega$, and amplitude $E_0$.   The hybridization function enters the KBE in the same manner as the self-energy, so the KBE in Eqs.~\ref{eq:KBEM}-\ref{eq:KBEL} are solved with $\Sigma\rightarrow\Sigma+\Delta$.  This introduces no algorithmic complications: in the example below, we simply define \texttt{DeltaPlusSigma} and compress the sum as a single object.
The impurity problem is solved using the fully self-consistent second Born approximation. In this approximation, the Fock term is given by
\begin{equation}
    \Sigma^F_{ij}(t) = -U(t) \rho_{ij}(t) \delta_{i\Bar{j}},
    \label{eq:SigmaF}
\end{equation} 
where $\Bar{i}=1-i$.
We work at half-filling, so that the Hartree term cancels the chemical potential.  For the first several iterations of the self-consistency loop in equilibrium, we apply a small source field $h^B_{ij}=\eta\delta_{i\bar{j}}$ to break symmetry and possibly induce a phase transition into the superconducting state.  
The time-dependent bubble and exchange diagrams are given by
\begin{align}
\begin{split}
    \Sigma^{B,\gtrless}_{ij}(t,t') &=   U(t) U(t') G^\gtrless_{ij}(t,t') G^\gtrless_{\Bar{i}\Bar{j}}(t,t') G^\lessgtr_{\Bar{j}\Bar{i}}(t',t),\\
    \Sigma^{E,\gtrless}_{ij}(t,t') &= - U(t) U(t') G^\gtrless_{\Bar{i}j}(t,t') G^\gtrless_{i\Bar{j}}(t,t') G^\lessgtr_{\Bar{j}\Bar{i}}(t',t).
    \label{eq:Sigma2B}
\end{split}
\end{align}
\ref{app:selfen} contains code implementing these expressions.

\subsection{Initialization of objects}
To begin the program, we define the scalar parameters for the calculation.  The number of timesteps \texttt{nt} and the number of HODLR levels \texttt{nlvl} define the geometry of the hierarchical decomposition, while \texttt{svdtol} is the truncation tolerance of the TSVD of each block.  The parameters \texttt{h} and \texttt{k} are the timestep size and integration order, respectively.  For systems initialized in thermal equilibrium, we must provide the inverse temperature \texttt{beta}, as well as the DLR parameters \texttt{epsdlr}, and \texttt{lambda} discussed in Sec.~\ref{ssec:DLR}.  The self-consistent iteration is controlled by the \texttt{MaxIter} and \texttt{MaxErr} parameters, which set the maximum number of iterations and the tolerance at which solutions are considered converged.  Lastly, the matrix dimension of the Green's function and self-energy is set by \texttt{nao}, and the particle statistics (+1 for bosons, -1 for fermions) is set by \texttt{xi}.
\begin{lstlisting}
// Hodlr parameters
int nlvl, nt, k;
double svdtol, h;

// Matsubara parameters
double epsdlr, lambda, beta;
int ntau;

// Self-Consistency Parameters
int BootMaxIter, StepMaxIter, MatsMaxIter;
double BootMaxErr, StepMaxErr, MatsMaxErr;

// Number of orbitals and particle statistics
int nao, xi;
\end{lstlisting}
These parameters can be read from an input file, which is provided as a command line argument:
\begin{lstlisting}
char* flin;
flin=argv[1];
find_param(flin,"__nlvl=",nlvl);
\end{lstlisting}
We then initialize the required objects using classes defined in \name, namely the DLR, integration weights, Dyson solver, Green's function, self-energy, self-energy evaluator, and quadratic Hamiltonian.  Note that the \texttt{dlr\_info} class takes the argument \texttt{ntau} by reference and updates it to the number of DLR nodes required to represent imaginary-time functions to the accuracy set by \texttt{lambda} and \texttt{epsdlr}.
The self-energy and hybridization functions, Eqs.~\ref{eq:Hybrid}, \ref{eq:SigmaF}, and \ref{eq:Sigma2B}, are implemented in the class \texttt{SC\_gf2}; a demonstration of their implementation is shown in \ref{app:selfen}. 
\begin{lstlisting}
// Dyson
hodlr::dlr_info dlr(ntau, lambda, epsdlr, beta, nao, xi);
Integration::Integrator I(k);
int r_v = RHO_DIAGONAL;
hodlr::dyson dyson_sol(nt, nao, k, dlr, r_v);
// Green's function storage
hodlr::herm_matrix_hodlr G(nt, ntau, nlvl, svdtol, nao, nao, xi, k);
hodlr::herm_matrix_hodlr DeltaPlusSigma(nt, ntau, nlvl, svdtol, nao, nao, xi, k);
// Self Energy evaluator
hodlr::SC_gf2 SC_gf2_solver(dlr);
// Time dependent Hamiltonian & chem. pot.
double mu = 0;
hodlr::function Hmf(nt, nao, nao);
hodlr::function U_func(nt, 1, 1);
hodlr::function t0_func(nt, nao, nao);
\end{lstlisting}

\subsection{Matsubara problem}
Once all objects are initialized, we solve the Matsubara problem self-consistently.  At each iteration, the Hartree-Fock self-energy, second-Born self-energy, and hybridization function are evaluated, and the Dyson equation is solved in the DLR basis \cite{kaye22,kaye23}.  The difference between the previous and updated $G^M$ is returned, and iterations continue until convergence.  After the Matsubara problem is solved, we must call the \texttt{initGMConvTensor} function of the Green's function object to calculate the imaginary-time convolution in the equation of the mixed component.  Note that the value of a $\texttt{function}$ object on the equilibrium branch $f(t=0_|)$ can be accessed by setting \texttt{tstp=-1}.
\begin{lstlisting}
// Do Matsubara iterations
double MatErr;
double eta = 0.0001;
int eta_steps = 5;
int tstp = -1;
for(int iter = 0; iter < MatsMaxIter; iter++) {
  // Reset Self-Energy
  DeltaPlusSigma.set_tstp_zero(-1);

  // Evaluate Delta via increment
  SC_gf2_solver.solve_Delta(tstp, t0_func, G, DeltaPlusSigma);

  // Evaluate Sigma via increment
  SC_gf2_solver.solve_Sigma(tstp, U_func, G, DeltaPlusSigma);

  // Evaluate Hmf
  SC_gf2_solver.solve_Sigma_Fock(tstp, U_func, G, Hmf);

  // For first several iters, add small field to break symmetry
  if(iter < eta_steps) {
    Hmf(tstp,0,1) += eta;
    Hmf(tstp,1,0) += eta;
  }

  // Solve Dyson Equation for G Matsubara
  MatErr = dyson_sol.dyson_mat(G, mu, Hmf, DeltaPlusSigma, false);

  // Exit if iteration difference is within tolerance.
  // Only exit after magnetic field is off
  if(MatErr < MatsMaxErr && iter > eta_steps) break;
}
// Matsubara is converged, we set the convolution tensor
G.initGMConvTensor(dlr);
\end{lstlisting}

\subsection{Bootstrapping}
\label{ssec:example_boot}
After solving the Matsubara problem, we begin the bootstrapping portion of the integration.  To obtain a starting point for the self-consistency loop, we evaluate the mean-field Green's functions as 
\begin{align}
    G^{R, HF}(t, t') &= -i \theta(t - t') \exp\left[ -i \epsilon(0_|) (t - t') \right]\\
    G^{<, HF}(t, t') &= i \exp\left[ -i \epsilon(0_|) t \right] \rho^M \exp\left[ i \epsilon(0_|) t' \right]\\
G^{\rceil, HF}(t, \tau) &= \xi i \exp\left[ -i \epsilon(0_|) t \right] \rho^M \exp\left[ \epsilon(0_|) \tau \right]
\end{align}
using the mean-field Matsubara Hamiltonian.  At each iteration, we first evaluate the self-energy and hybridization, Eqs.~\ref{eq:Hybrid}, \ref{eq:SigmaF}, and \ref{eq:Sigma2B}, for $0\leq t\leq k$ (with $k$ the order of accuracy), before solving the KBE in this same region.  Just as in the Matsubara case, the \texttt{dyson} integrator returns the difference between the previous and current iterations of $G$, which is used to monitor convergence.

\begin{lstlisting}
//Self-consistency for bootstrapping
dyson_sol.green_from_H(G, mu, Hmf, h, k);
for(int iter = 0; iter < BootMaxIter; iter++) {

  // Evaluate Hmf
  for(int tstp = 0; tstp <= k; tstp++) {
    SC_gf2_solver.solve_Sigma_Fock(tstp, U_func, G, Hmf);
  }

  // Evaluate Self-Energy 
  for(int tstp = 0; tstp <= k; tstp++) {
    DeltaPlusSigma.set_tstp_zero(tstp);
    SC_gf2_solver.solve_Delta(tstp, t0_func, G, DeltaPlusSigma);
    SC_gf2_solver.solve_Sigma(tstp, U_func, G, DeltaPlusSigma);
  }

  // Solve Dyson
  double err = dyson_sol.dyson_start_ntti(G, mu, Hmf, DeltaPlusSigma, I, h);

  if(err < BootMaxErr) break;
}
\end{lstlisting}

\subsection{Timestepping and SVD updates}
After the bootstrapping procedure has converged, we proceed to timestepping.  To begin each step, we update the HODLR representations of the Green's function and self-energy.  We then extrapolate the Green's function to get an initial guess for the self-consistent iteration.  Within the self-consistency loop, we first evaluate the mean-field and correlated self-energies before calling the timestep function provided by the \texttt{dyson} class, which again returns the difference between $G$ and its previous iteration.
\begin{lstlisting}
for(int tstp = k+1; tstp < nt; tstp++) {
  G.update_blocks(I);
  DeltaPlusSigma.update_blocks(I);

  dyson_sol.Extrapolate(G, I);

  for(int iter = 0;  iter < StepMaxIter; iter++) {
    // HF
    SC_gf2_solver.solve_Sigma_Fock(tstp, U_func, G, Hmf);

    // 2B
    DeltaPlusSigma.set_tstp_zero(tstp);
    SC_gf2_solver.solve_Delta(tstp, t0_func, G, DeltaPlusSigma);
    SC_gf2_solver.solve_Sigma(tstp, U_func, G, DeltaPlusSigma);

    // Dyson
    double err = dyson_sol.dyson_timestep(tstp, G, mu, Hmf, DeltaPlusSigma, I, h);

    if(err < CorrectorMaxErr) break;
  }
}
\end{lstlisting}

\subsection{Outputting results}
Once the calculations are complete, the data may be printed to an \texttt{hdf5} file using the following functions.  A python notebook is included in the \texttt{Plotting} directory which reads in $G^R$ and $G^<$ printed using the \texttt{write\_to\_hdf5} function.
\begin{lstlisting}
// Open output file
h5e::File out_file("out.hdf5", h5e::File::Overwrite | h5e::File::ReadWrite | h5e::File::Create);
// Output for Green's function and self-energy
G.write_to_hdf5(out_file, "G");
DeltaPlusSigma.write_to_hdf5(out_file, "S");
// Output single-time functions
Hmf.write_to_hdf5(out_file, "H");
// Write timing information is profiling if enabled
dyson_sol.write_timing(out_file, "dyson");
\end{lstlisting}

\subsection{Checkpoint files}
Calculations that must be interrupted and restarted can make use of \texttt{hdf5} checkpoint files, which store the full simulation state.  These files can initialize objects via specialized constructors, and are supported for the \texttt{herm\_matrix\_hodlr}, \texttt{function}, and \texttt{dyson} classes.  An example of writing a checkpoint file is as follows:
\begin{lstlisting}
// Open checkpoint file
h5e::File check_file("check.hdf5", h5e::File::Overwrite | h5e::File::ReadWrite | h5e::File::Create);
// Checkpoints for Green's function and self-energy
G.write_checkpoint_hdf5(check_file, "G");
DeltaPlusSigma.write_checkpoint_hdf5(check_file, "S");
// Checkpoints for single-time functions
Hmf.write_checkpoint_hdf5(check_file, "H");
// Checkpoints for dyson integrator only necessary for timing information
dyson_sol.write_checkpoint_hdf5(check_file, "dyson");
\end{lstlisting}
A calculation can then be restarted by reading from this file:
\begin{lstlisting}
h5e::File check_file("check.hdf5", h5e::File::ReadOnly);
hodlr::herm_matrix_hodlr G(check_file, "G");
hodlr::herm_matrix_hodlr DeltaPlusSigma(check_file, "S");
hodlr::function Hmf(check_file, "H");
hodlr::dyson dyson_sol(check_file, "dyson", dlr);
\end{lstlisting}

\subsection{Timing results}
\begin{figure}
    \centering
    \includegraphics[width=0.95\linewidth]{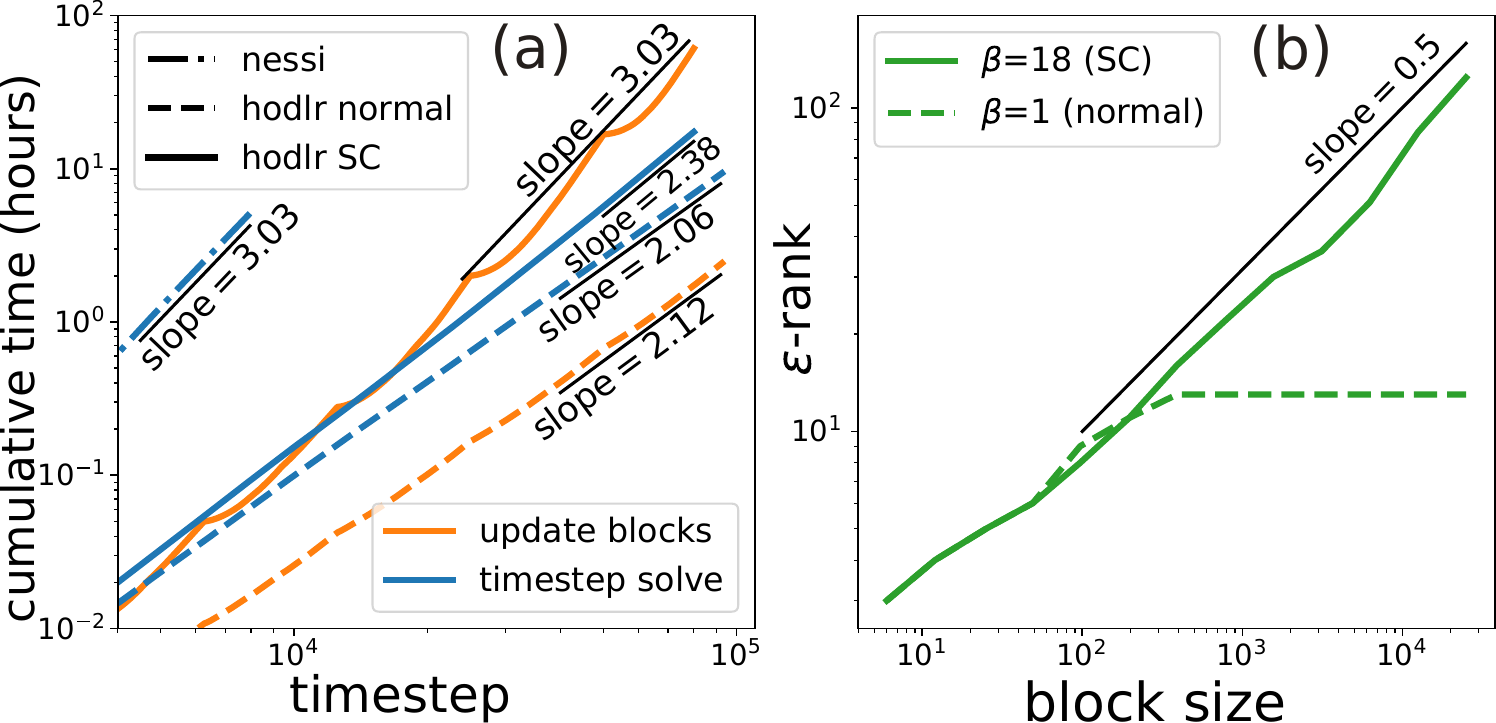}
    \caption{(a)~Wall clock time for superconducting~(solid lines) and disordered (dashed lines) attractive Hubbard model with asymptotic scaling fits~(black lines).  Blue and orange lines show contribution of solving timesteps and updating SVD blocks, respectively.  The timing of the NESSi solver~\cite{schuler2020} (dot-dashed line) is shown for comparison. (b) Scaling of the $\varepsilon$-ranks versus the block size in the superconducting state at $\beta=18$~(SC) and the disordered state at $\beta=1$~(normal).  Hamiltonian parameters are $U=2$, $E_0=0.02$, $\sigma=6.5$, $t_c=16$, $\omega=1.3$.  H-NESSi data obtained using $\epsilon_{SVD}=10^{-6}$.}
    \label{fig:wall_vs_tstp}
\end{figure}

To demonstrate the efficiency of the \name{} code, we plot the wall clock time required to reach a given timestep and the $\varepsilon$-rank for growing block sizes in Figs.~\ref{fig:wall_vs_tstp}(a) and (b), respectively.  We compare three different calculations: a reference implementation in the NESSi package, a calculation from \name{} in the superconducting state, and a calculation from \name{} in the normal state.  From (b), we can see that the $\varepsilon$-rank of the Green's function has very different behavior in the two different phases.  The normal state is highly compressible, and the ranks of the blocks saturate at small block sizes, whereas in the superconducting state, the ranks grow as $N^{1/2}$, where $N$ is the size of the block.  The behavior of the ranks affects the efficiency of our integration algorithm, as shown in (a).  For the Dyson solver, both the superconducting and normal states scale approximately as $T^{2+\alpha}$, where $\alpha$ is the exponent characterizing the growth of the $\epsilon$-rank with block size (e.g., $\alpha \approx 0.5$ for the superconducting state and $\alpha \approx 0$ for the normal state).  This is significantly better than the $T^{3}$ scaling of the NESSi solver (dot-dashed line in (a)).  Furthermore, the scaling prefactors are observed to be significantly smaller in \name{}.  We also show timings for the TSVD update step in orange.  The observed scaling matches the expected scaling of $T^{2+2\alpha}$ for both the normal and superconducting phases.

\section{MPI parallelization for generic lattice}
Many systems of interest possess translational invariance, so that the Green's functions $G_{\mathbf{k},ij}$ and self-energies $\Sigma_{\mathbf{k},ij}$ are block-diagonal in momentum $\mathbf{k}$, with $ij$ indexing non-diagonalizable orbital degrees of freedom.  The Dyson equation then decouples across $\mathbf{k}$-points, enabling straightforward parallelization.  However, the self-energy is typically a functional of the Green's function over the entire Brillouin zone, not just at the local $\mathbf{k}$-point.  In a distributed-memory setting, inter-process communication of $G_{\mathbf{k},ij}$ is therefore required so that each process can evaluate $\Sigma_{\mathbf{k},ij}$ for its assigned $\mathbf{k}$-points.

\begin{figure*}
	\includegraphics[width=\textwidth]{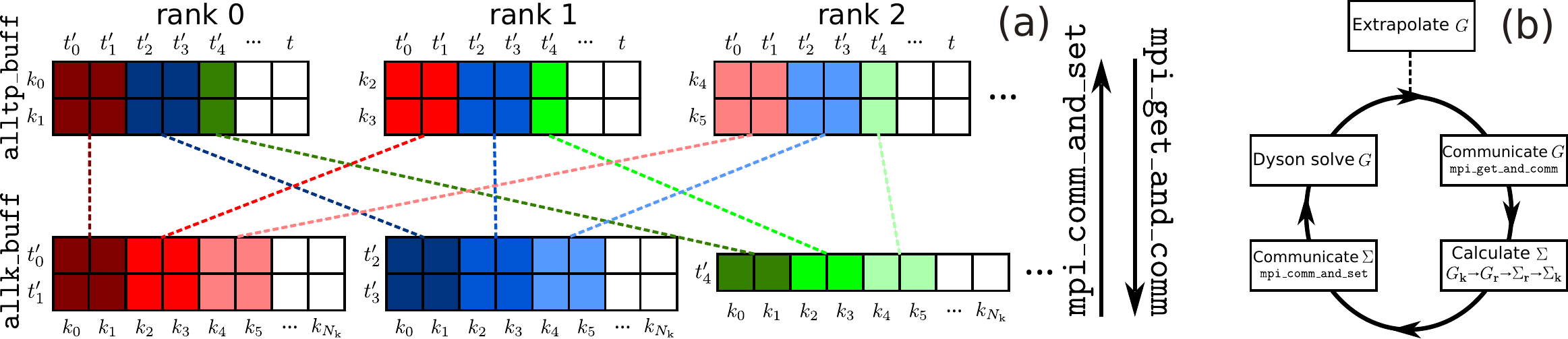}
	\centering
	\caption{(a)~Schematic of the MPI communication performed by the \texttt{mpi\_comm} class between the \texttt{alltp\_buff} and \texttt{allk\_buff} data buffers; see the main text for details.  Squares of the same color represent the same data, and colored dashed lines indicate their distribution between the two buffers.  White squares denote allocated buffer space for future time points or additional momentum points beyond the active block. For the specific parameters shown ($N_k=6$, $N_{\mathrm{mpi}}=3$, and communication up to $t'_4$), the colored blocks account for the entirety of the actively communicated data, illustrating the complete data transposition. (b)~Typical self-consistency loop for lattice calculations.}
	\label{fig:communication}
\end{figure*}

This section is organized as follows.  Section~\ref{sec:mpi_comm} introduces the \texttt{mpi\_comm} class, which manages internode communication with minimal user effort and an interface mirroring that of \texttt{herm\_matrix\_hodlr}.  Section~\ref{sec:driven2d_example} presents example calculations for the two-dimensional Hubbard model using a massively parallel implementation, comparing performance with NESSi and analyzing Green's function compression.  Section~\ref{sec:driven2d_implementation} provides implementation details of our hybrid MPI + OpenMP parallelization, and Section~\ref{sec:driven2d_performance} discusses scaling behavior.

\subsection{The {\tt{mpi\_comm}} class}\label{sec:mpi_comm}

The \texttt{mpi\_comm} class in \name{} manages workload distribution across MPI processes, communication of Green's function and self-energy data, and intraprocess parallelization via OpenMP threading.  Upon initialization, the \texttt{mpi\_comm} object exposes \texttt{my\_Nk}, the number of Brillouin zone points assigned to the local process, and \texttt{my\_global\_k\_list}, the corresponding global indices.  These quantities are then used to initialize the Green's functions, self-energies, and any auxiliary containers such as a \texttt{function} object storing the $\mathbf{k}$-dependent Hamiltonian:
\begin{lstlisting}
mpi_comm comm(global_Nk, Nt, r, nao);
int my_Nk = comm.my_Nk;
std::vector<herm_matrix_hodlr> G_vec;
std::vector<herm_matrix_hodlr> S_vec;
std::vector<function> H_vec;
G_vec.reserve(my_Nk);
S_vec.reserve(my_Nk);
H_vec.reserve(my_Nk);
for(int i=0;i<my_Nk;i++){
    G_vec.emplace_back(Nt,r,nlvl,svdtol,nao,nao,xi,SolverOrder);
    S_vec.emplace_back(Nt,r,nlvl,svdtol,nao,nao,xi,SolverOrder);
    H_vec.emplace_back(Nt,nao,nao);
}
\end{lstlisting}

At each timestep, the self-consistency loop shown in Fig.~\ref{fig:communication} iterates the self-energy and Green's function until convergence.  Evaluating the self-energy at a given $\mathbf{k}$-point requires knowledge of $G_{\mathbf{k}'}$ for all $\mathbf{k}'$ in the Brillouin zone.  This communication is handled by two functions: \texttt{mpi\_get\_and\_comm}, which loads a vector of the locally stored Green's functions into MPI communication buffers and distributes them to all processes, and \texttt{mpi\_comm\_and\_set}, which communicates the evaluated self-energies back and extracts them into a vector of locally-stored self-energies.  For both functions, only the Matsubara component is communicated when \texttt{tstp==-1}; otherwise the lesser, retarded, and mixed components at the current timestep are all communicated.  For imaginary-time functions, the ``reversed'' values $G^M(\beta-\tau_i)$ and $G^\rceil(t,\beta-\tau_i)$ are also communicated, as they are often needed in the self-energy evaluation.  Each communication function comes in two variants: a \texttt{\_spawn} variant, which creates OpenMP threads at each call via \texttt{\#pragma omp parallel} to accelerate buffer loading and unloading, and a \texttt{\_nospawn} variant, which assumes it is called from within an existing OpenMP parallel region.

The left-hand side of Fig.~\ref{fig:communication} illustrates the communication scheme.  Before communication, each rank stores all time points (and DLR nodes, although only the real-time buffers are depicted) at the current timestep for its locally assigned $\mathbf{k}$-points, in the buffer \texttt{alltp\_buff}.  After communication, each rank instead holds every $\mathbf{k}$-point but only a subset of time points, in the buffer \texttt{allk\_buff}.  In the illustrated example, at time $t=4$, rank~1 receives $G^R_\mathbf{k}(t=4,t')$ and $G^<_\mathbf{k}(t',t=4)$ for $t'=2,3$ and every $\mathbf{k}$.  Each rank is then responsible for evaluating the self-energy at its assigned subset of time points (and DLR nodes, for the mixed component) using the communicated Green's function data.  The number of assigned time points is given by \texttt{my\_Nt}, and the first assigned time point by \texttt{my\_first\_t}.  The mixed and Matsubara components are split analogously, with \texttt{my\_Ntau} and \texttt{my\_first\_tau} specifying the local imaginary-time nodes.  The communicated data is accessed through an interface that returns \texttt{Eigen} matrix maps of the internal buffers; these functions, listed in Table~\ref{tab:comm_maps}, closely mirror those of the \texttt{herm\_matrix\_hodlr} class.
\begin{table}[h!]
    \centering
    \begin{tabular}{|c|c|}\hline
        \texttt{map\_ret} & \makecell{Return $N_o\times N_o$ \texttt{Eigen} matrix map \\of $C^R_{ij}(t,t')$ at given $t'$ and $\mathbf{k}$\\for communicated timestep $t$} \\\hline
        \texttt{map\_les} & \makecell{Return $N_o\times N_o$ \texttt{Eigen} matrix map \\of $C^<_{ij}(t',t)$ at given $t'$ and $\mathbf{k}$\\for communicated timestep $t$} \\\hline
        \texttt{map\_tv} & \makecell{Return $N_o\times N_o$ \texttt{Eigen} matrix map \\of $C^\rceil_{ij}(t,\tau)$ at given DLR node $\tau_i$ and $\mathbf{k}$\\for communicated timestep $t$} \\\hline
        \texttt{map\_tv\_rev} & \makecell{Return $N_o\times N_o$ \texttt{Eigen} matrix map \\of $C^\rceil_{ij}(t,\beta-\tau)$ at given DLR node $\tau_i$ and $\mathbf{k}$\\for communicated timestep $t$} \\\hline
        \texttt{map\_mat} & \makecell{Return $N_o\times N_o$ \texttt{Eigen} matrix map \\of $C^M_{ij}(\tau)$ at given DLR node $\tau_i$ and $\mathbf{k}$} \\\hline
        \texttt{map\_mat\_rev} & \makecell{Return $N_o\times N_o$ \texttt{Eigen} matrix map \\of $C^M_{ij}(\beta-\tau)$ at given DLR node $\tau_i$ and $\mathbf{k}$} \\\hline
    \end{tabular}
    \caption{Functions for reading and writing to the communication buffers internal to \texttt{mpi\_comm}.}
    \label{tab:comm_maps}
\end{table}

Since the \texttt{mpi\_comm} class handles all communication and buffer management, the user's only responsibility is to evaluate the self-energy $\Sigma_\mathbf{k}$ from $G_\mathbf{k}$.  In practice, this amounts to using the map functions in Table~\ref{tab:comm_maps} to \textit{replace} the communicated $G_\mathbf{k}$ data with the corresponding $\Sigma_\mathbf{k}$ data.  A typical implementation has the following structure:
\begin{lstlisting}
// Communicate G
comm.mpi_get_and_comm_spawn(tstp, G_vec, dlr);
// mpi_comm assigns each mpi rank a range for t'
int init_t = comm.my_first_t;
int last_t = init_t + comm.my_Nt;
// each rank must do assigned t' and all k
for(int t = init_t; t < last_t; t++) {
    for(int k = 0; k < global_Nk; k++) {
        // Read in G ret,les before overwriting
        ZMatrix GR = comm.map_ret(k,t);
        ZMatrix GL = comm.map_les(k,t);
        // Evaluate Sigma into comm buffers
        comm.map_ret(k,t) = ...
        comm.map_les(k,t) = ...
    }
}
// mpi_comm assigned range for dlr points
int init_tau = comm.my_first_tau;
int last_tau = init_tau + comm.my_Ntau;
// each rank must do assigned tau and all k
for(int tau = init_tau; tau < last_tau; tau++) {
    for(int k = 0; k < global_Nk; k++) {
        // Read in G mix before overwriting
        ZMatrix GTV = comm.map_tv(k,tau);
        // Evaluate Sigma into comm buffers
        comm.map_tv(k,tau) = ...
    }
}
// Communicate Sigma
comm.mpi_comm_and_set_spawn(tstp, S_vec);
\end{lstlisting}

This default interface should suffice for nearly all use cases.  For situations requiring communication of additional Keldysh components or only a subset of orbital indices, a low-level interface to the underlying communication buffers is described in \ref{app:mpi_comm_interface}.

\subsection{Driven 2D Hubbard example}
\label{sec:driven2d_example}
We demonstrate the parallel implementation by computing the optical conductivity of the Hubbard model.  The system is probed with a short electric field pulse, and the optical conductivity is extracted by inverting the linear-response relation. The pulse preserves translational invariance, enabling large-scale parallelization of the \texttt{Dyson} solver.  Such parallelization is essential: the lattice must be large enough to suppress finite-size effects and the propagation long enough for the current response to fully decay, so that in physically relevant parameter regimes RAM usage easily exceeds $10$~TB even with HODLR compression. This example is therefore well suited to illustrate how hierarchical compression enables otherwise infeasible calculations. We require a highly parallel implementation for this example: for the largest lattices and lowest temperatures considered here, we require more than $15$ thousand CPU cores across $60$ nodes.
\begin{figure}
\includegraphics[width = 0.4\textwidth]{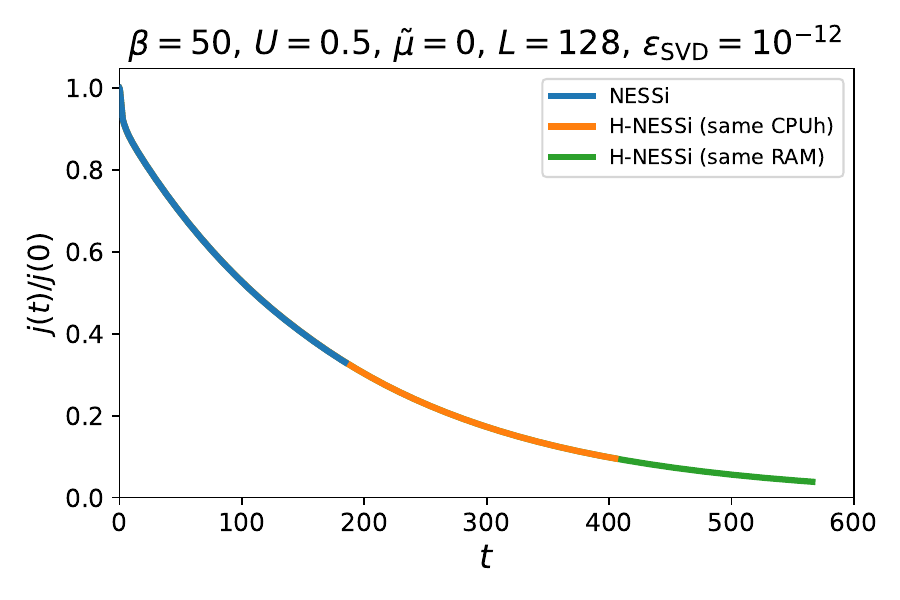}
\centering
\caption{Current obtained from the solution of the KB equations via Eq.~\ref{eq:current}.  The NESSi calculation is limited to $N_\mathrm{t}=2048$ time points, whereas \name{} reaches $N_\mathrm{t}=6289$ time points under the same memory constraints (green curve).  For the same number of CPU hours, \name{} propagates to longer physical times, as shown by the orange curve.}
\label{fig:currents}
\end{figure}

Below, we summarize the physical setup, then outline the hybrid MPI + OpenMP parallelization scheme, which combines our distributed data containers with efficient global communication and shared-memory parallelism on each MPI rank.  Finally, we compare memory usage and runtimes with NESSi and study the scaling behavior of the algorithm.

We consider the square lattice Hubbard model, described by the Hamiltonian
\begin{equation}\label{eq:hamiltonian}
	H = -J\sum_{\langle ij \rangle,\sigma} c^\dagger_{\sigma,i}c_{\sigma,j} -\mu \sum_{\sigma, i} n_{\sigma,i} + U\sum_i n_{\uparrow,i}n_{\downarrow,i}
\end{equation}
where $i,j$ enumerate lattice sites, $c^\dagger$ and $c$ are creation and annihilation operators, $\sigma \in \{\uparrow,\downarrow\}$ denotes spin, $J$ is the nearest-neighbor hopping amplitude, $n_{\sigma,i}$ is the particle-number operator, $\mu$ is the chemical potential, and $U$ is the on-site interaction strength.  In practice, the Hartree shift is absorbed into the chemical potential, $\tilde{\mu} = \mu - U\langle n_{\sigma,i} \rangle$.  The system is probed by a spatially uniform electric field in the $x$-direction, introduced through the vector potential $A(t)$ via the Peierls substitution $\varepsilon_{\mathbf{k}} \to \varepsilon_{\mathbf{k}-\mathbf{A}(t)}$, where $\varepsilon_\mathbf{k}$ is the non-interacting dispersion relation.  For a given self-energy approximation, we solve the KB equations for the Green's function and compute the current, kinetic energy, and average occupancy per spin as
\begin{align}
	\langle j(t) \rangle &= -\frac{2i}{N_\mathrm{k}}\sum_{\mathbf{k}} v^x_{\mathbf{k}-\mathbf{A}(t)} G^<_{\mathbf{k}}(t,t)\label{eq:current}\\
    E_\mathrm{kin}&=-2i/N_{k}\sum_{\mathbf{k}}\varepsilon_{\mathbf{k}}G^<_{\mathbf{k}}(t,t)\\
    \langle n \rangle&=-i/N_{k}\sum_{\mathbf{k}}G^<_{\mathbf{k}}(t,t),
\end{align}
where $N_\mathrm{k} = L^2$ is the number of $\mathbf{k}$ points, $L$ is the linear lattice size, and $\mathbf{v} = \nabla_{\mathbf{k}}\varepsilon_{\mathbf{k}-\mathbf{A}(t)}$ is the time-dependent velocity. 
We choose a step-function vector potential,
\begin{equation}
	A(t) = -A_\mathrm{max} \theta(t) \implies E(t) = A_\mathrm{max} \delta(t)
\end{equation}
where $A_{\mathrm{max}}$ is the amplitude of the abrupt shift at $t=0$ and $E(t) = -\partial_t A(t)$.
The $\delta$-function form of the electric field allows us to invert the linear-response relation,
\begin{equation}
	\langle j(t) \rangle = \int_{0}^{t}dt' \sigma(t-t') E(t') \implies \sigma(t) = \frac{\langle j(t)\rangle}{A_\mathrm{max}},
\end{equation}
and extract the optical conductivity $\sigma(t)$ directly from the computed current.
For a short pulse of the electric field in the $x$-direction, the Green's functions and the self-energy have the reflection symmetry
\begin{equation}
	G_{(k_x,k_y)}(t,t') = G_{(k_x,-k_y)}(t,t'),\label{eq:reflection_symm}
\end{equation}
allowing us to consider a smaller number of $\mathbf{k}$-points, 
$L^2 \to L\left(\frac{L}{2}+1\right)$.
\begin{figure}
	\includegraphics[scale=0.5]{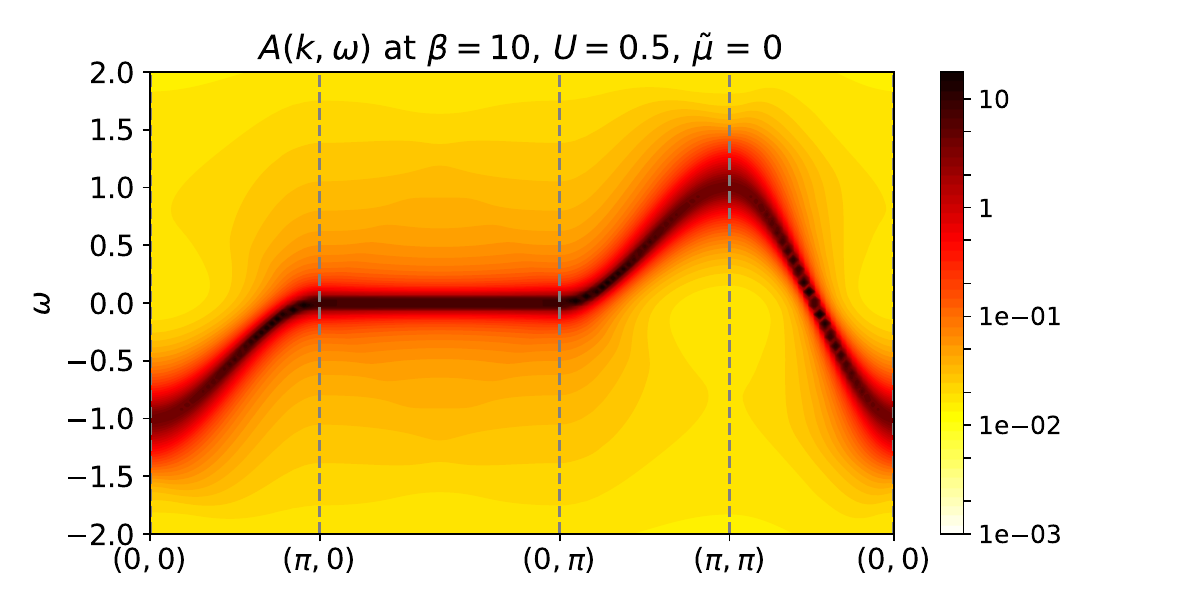}
	\caption{Equilibrium spectral function $A(k,\omega)$ for the Hubbard Hamiltonian at half-filling, with $U=0.5$ and $\beta=10$.}
	\label{fig:dos}
\end{figure}

\begin{figure*}
	\includegraphics[width=1\textwidth]{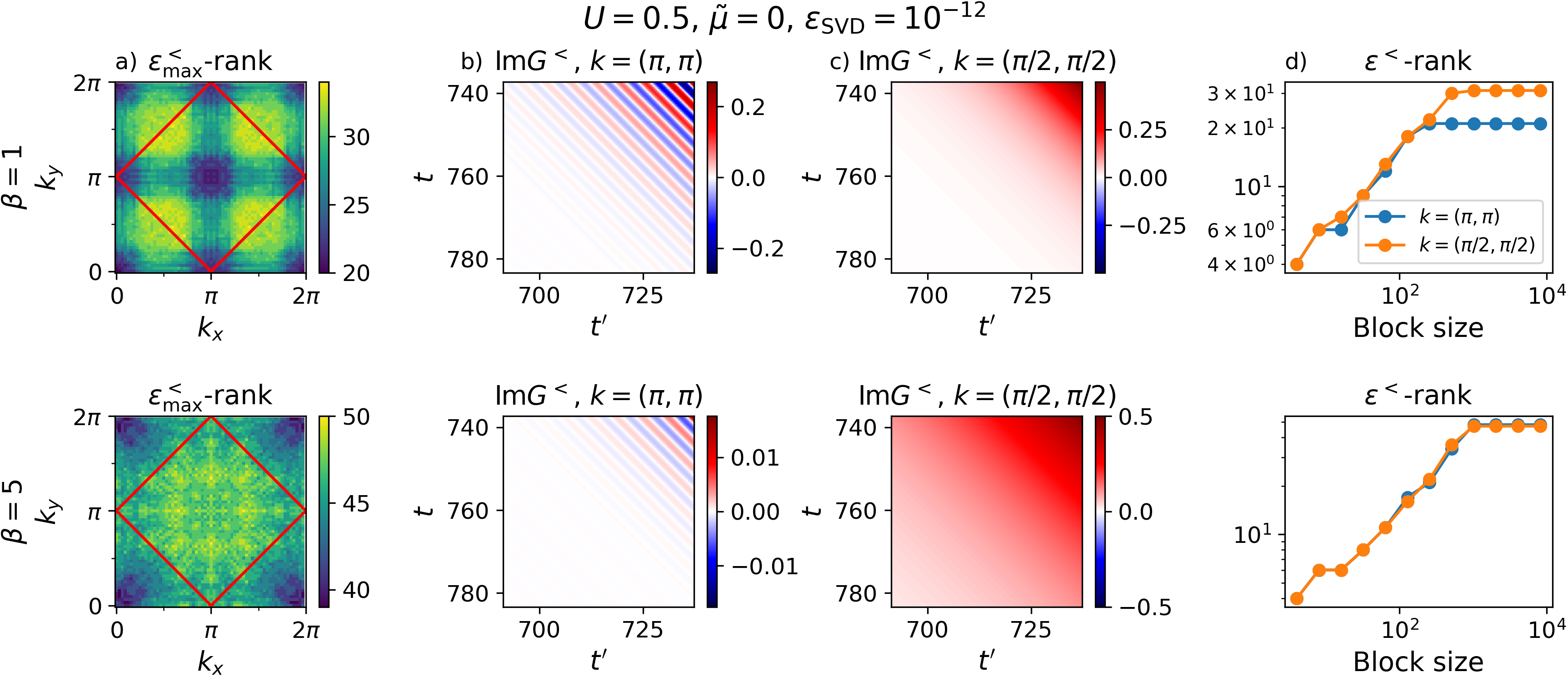}
	\centering
	\caption{(a)~Maximum $\varepsilon$-rank across different $\mathbf{k}$-points. The red line indicates the Fermi surface.  (b,\,c)~Largest block of the imaginary part of the lesser Green's function $G^<$ at $\mathbf{k} = (\pi,\pi)$ and $\mathbf{k} = (\pi/2,\pi/2)$.  (d)~$\varepsilon$-rank of $G^<$ as a function of block size.  All results are shown for two temperatures, with lattice size $L=64$ and $N_t=16384$ timesteps.}
	\label{fig:ranks}
\end{figure*}

To close the system of KB equations, we approximate the self-energy with the self-consistent Born approximation, which for the Hubbard Hamiltonian reads
\begin{align}
	\Sigma^M_{\mathbf{r}}(\tau) &= U^2 G^M_{\mathbf{r}}(\tau)G^M_{\mathbf{r}}(\tau)G^M_{-\mathbf{r}}(-\tau)\label{eq:BornApproxM}\\
	\Sigma^\gtrless_{\mathbf{r}}(t,t') &= U^2 G^\gtrless_{\mathbf{r}}(t,t')G^\gtrless_{\mathbf{r}}(t,t')G^\lessgtr_{-\mathbf{r}}(t',t)\label{eq:BornApproxLG}\\
	\Sigma^{\lceil}_{\mathbf{r}}(t,\tau) &= U^2 G^\lceil_{\mathbf{r}}(t,\tau)G^\lceil_{\mathbf{r}}(t,\tau)G^\rceil_{-\mathbf{r}}(\tau,t),\label{eq:BornApproxtv}	
\end{align}
where $\mathbf{r}$ denotes real-space lattice sites.  We note that these expressions are much more expensive to evaluate when written in the momentum basis. We therefore use \texttt{FFTW}~\cite{FFTW.jl-2005} to Fourier transform $G_\mathbf{k}\rightarrow G_\mathbf{r}$ and inverse-transform $\Sigma_\mathbf{r}\rightarrow \Sigma_\mathbf{k}$.  Performing these transforms requires every $\mathbf{k}$-point on each MPI process, which is precisely the communication protocol implemented in the \texttt{mpi\_comm} class.

Figure~\ref{fig:currents} shows an example current obtained with both NESSi and \name{}.  Memory constraints limit the NESSi calculation to $N_\mathrm{t}=2048$ time points, whereas \name{} reaches $N_\mathrm{t}=6289$ time points with the same memory budget (green curve in Fig.~\ref{fig:currents}).  For an equal number of CPU hours, \name{} propagates to longer physical times than NESSi (orange curve).  In this example, \name{} reaches a propagation time roughly three times larger than NESSi under identical memory constraints. For larger $N_{\mathrm{t}}$ the advantage is even more pronounced, reflecting the improved memory scaling: $\mathcal{O}(N_\mathrm{t}^2)$ for NESSi versus $\mathcal{O}(N_\mathrm{t}N_S^\varepsilon)$ for \name{}.
At lower temperatures the current decays substantially more slowly, requiring longer propagation times. Simultaneously, larger lattices are needed to suppress finite-size effects.  Both factors increase memory requirements and make the calculations more demanding.


Within this setup, equilibrium single-particle correlators in the real-time domain can be obtained by simply setting the external field to zero.
The self-consistent Born approximation is less expensive when formulated in real time, scaling as $\mathcal{O}(N_t)$, than in real frequency, where a straightforward implementation scales as $\mathcal{O}(N_\omega^2)$.  Although the Dyson solver is typically more expensive in real time, the overall calculation remains inexpensive and is straightforward to set up with our code.  As an example, the equilibrium spectral function is shown in Fig.~\ref{fig:dos}.

Figure~\ref{fig:ranks}(a) shows that the compressibility of the Green's function varies significantly across $\mathbf{k}$-points: the least compressible points require nearly twice as many singular values as the most compressible ones.  At lower temperatures, this spread appears to narrow, but the overall number of required singular values increases, reflecting the richer structure of low-temperature correlators.  Despite these differences, the $\epsilon$-rank at all $\mathbf{k}$-points saturates at relatively small block sizes, consistent with optimal quadratic scaling of the integration routines.  At block sizes of $8129$, the memory savings relative to dense storage reach nearly a factor of $100$ at $\beta=5$.

\subsection{Implementation details}\label{sec:driven2d_implementation}
The computation of the self-energy is made drastically cheaper by translation symmetry (even in the presence of a uniform electric field encoded in the time-dependent vector potential).  As seen from Eqs.~\ref{eq:BornApproxM}--\ref{eq:BornApproxtv}, the self-energy is diagonal in the position basis, unlike in the momentum representation.  Moreover, the expression is time-local, enabling efficient parallelization through the communication protocol of \texttt{mpi\_comm} described in Section~\ref{sec:mpi_comm}.

The self-energy is most efficiently evaluated in real space, whereas the KB equations decouple in the $\mathbf{k}$-space representation.  We bridge the two domains as follows:
\begin{enumerate}
    \item \emph{Distribution of Green's functions}:
    \texttt{mpi\_comm} distributes the Green's functions via \texttt{mpi\_get\_and\_comm}, providing each MPI process with all $\mathbf{k}$-points for its assigned range of time points.
    \item \emph{Real-space self-energy evaluation}:
    Each MPI process loops over its assigned time (and DLR) points and performs the following steps:
    \begin{itemize}
        \item Inverse 2D Fourier transform $G_\mathbf{k}\rightarrow G_\mathbf{r}$ using \texttt{FFTW}.
        \item Evaluate the self-energy in the real-space basis via Eqs.~\ref{eq:BornApproxM}--\ref{eq:BornApproxtv}.
        \item Forward 2D Fourier transform $\Sigma_\mathbf{r}\rightarrow \Sigma_\mathbf{k}$ using \texttt{FFTW}.
    \end{itemize}
    Since these operations are independent at each time (and DLR) point, they can be parallelized over threads with OpenMP.
    \item \emph{Distribution of self-energies}:
    \texttt{mpi\_comm} redistributes the self-energies via \texttt{mpi\_comm\_and\_set}, returning each process's local $\mathbf{k}$-points.
\end{enumerate}
This self-energy evaluation scheme follows exactly the example implementation in Sec.~\ref{sec:mpi_comm}.
The most expensive step in the procedure (aside from internode communication) is the Fourier transforms, which scale as $\mathcal{O}\left(N_\mathrm{k}\log N_\mathrm{k}\right)$, compared to the $\mathcal{O}\left(N_{\mathrm{k}}^2\right)$ cost of evaluating the self-energy directly in momentum space.
This communication and evaluation process is encapsulated in the \texttt{born\_approx\_se} class, which also exploits the time-locality of the self-energy to parallelize over time and DLR points using OpenMP:
\begin{lstlisting}
born_approx_se se_eval(U,L,global_Nk,size,nthreads);
\end{lstlisting}

The \texttt{lattice} object stores information about all global $\mathbf{k}$-points in the Brillouin zone, accounting for the reflection symmetry of Eq.~\ref{eq:reflection_symm}, and provides the time-dependent non-interacting dispersion relation $\varepsilon_{\mathbf{k}}(t)$:
\begin{lstlisting}
lattice_2d_ysymm lattice(L, Nt, dt, J, Amax, mu, nao);
\end{lstlisting}

The Dyson solver is parallelized over local $\mathbf{k}$-points using OpenMP.  Because the \texttt{dyson} class uses internal buffers, it is not thread-safe; each thread must therefore have its own solver instance.  The same applies to the \texttt{dlr\_info} class, whose off-node evaluation functions (Table~\ref{tab:getmat_offnode}) are likewise not thread-safe:
\begin{lstlisting}
//dlr vector
std::vector<hodlr::dlr_info> dlr_vec;
dlr_vec.reserve(nthreads);
for(int i = 0; i < nthreads; i++) dlr_vec.emplace_back
  (r, dlrlambda, epsdlr, beta, nao, xi);

std::vector<hodlr::dyson> dyson_sol_vec;
dyson_sol_vec.reserve(nthreads);
  for(int i=0; i<nthreads; i++) dyson_sol_vec.emplace_back(Nt, nao, SolverOrder, dlr_vec[i], rho_version, profile);
\end{lstlisting}

Each $\mathbf{k}$-point is represented by an instance of the \texttt{kpoint} class, which stores two \texttt{herm\_matrix\_hodlr} objects for the Green's function and self-energy, the non-interacting dispersion $\varepsilon_{\mathbf{k}}(t)$, and methods that perform the Dyson propagation at that $\mathbf{k}$-point.  Each MPI rank holds its assigned $\mathbf{k}$-points in a vector \texttt{corrK\_rank} of \texttt{kpoint} objects:
\begin{lstlisting}
//kpoint contains Gs and Sigmas
std::vector<std::unique_ptr<kpoint>> corrK_rank;
corrK_rank.resize(my_Nk);

//create kpoints either from scratch or from checkpoint
if(!checkpoint_exists){
  for(int i=0;i<my_Nk;i++){
    int global_k = comm.my_global_k_list[i];
    corrK_rank[i] = std::make_unique<kpoint>(Nt,r,nlvl,svdtol,nao,beta,dt,SolverOrder,lattice.kpoints_[global_k],lattice,mu,Ainitx);
  }
}
else{
  for(int i=0;i<my_Nk;i++){
    int global_k = comm.my_global_k_list[i];
    h5e::File checkpoint_file(checkpoint_dir+"GSigma" + std::to_string(i)+".h5", h5e::File::ReadOnly);
    corrK_rank[i] = std::make_unique<kpoint>(Nt,r,nlvl,svdtol,nao,beta,dt,SolverOrder,lattice.kpoints_[global_k],lattice,mu,Ainitx,checkpoint_file);      
  }
}
\end{lstlisting}
Observables are computed on each MPI rank for their local $\mathbf{k}$-blocks during timestepping and accumulated across all ranks via \texttt{MPI\_Reduce} at the end of the calculation.

The Dyson integration is carried out by the \texttt{step\_dyson} method of the \texttt{kpoint} class, which calls \texttt{dyson\_mat}, \texttt{dyson\_start\_ntti}, or \texttt{dyson\_timestep} for the Matsubara, bootstrapping, and timestepping stages, respectively.  Since each $\mathbf{k}$-point is independent, this loop is parallelized with OpenMP; as noted above, each thread must use its own \texttt{dyson} and \texttt{dlr\_info} instances:
\begin{lstlisting}
std::vector<double> errk(my_Nk);
#pragma omp parallel
{
  int thread_id = omp_get_thread_num();
  #pragma omp for schedule(static)
  for(int k=0;k<my_Nk;k++){
    errk[k] = corrK_rank[k]->step_dyson(tstp, SolverOrder, lattice, I, dyson_sol_vec[thread_id], dlr_vec[thread_id]);
  }
}
// gather all errors from each k point
double err;
double tot_err;
err = std::reduce(errk.begin(),errk.end()); 
MPI_Allreduce(&err, &tot_err, 1, MPI_DOUBLE, MPI_SUM, MPI_COMM_WORLD);
\end{lstlisting}

\begin{figure*}[!h]
	\includegraphics[width=0.85\textwidth]{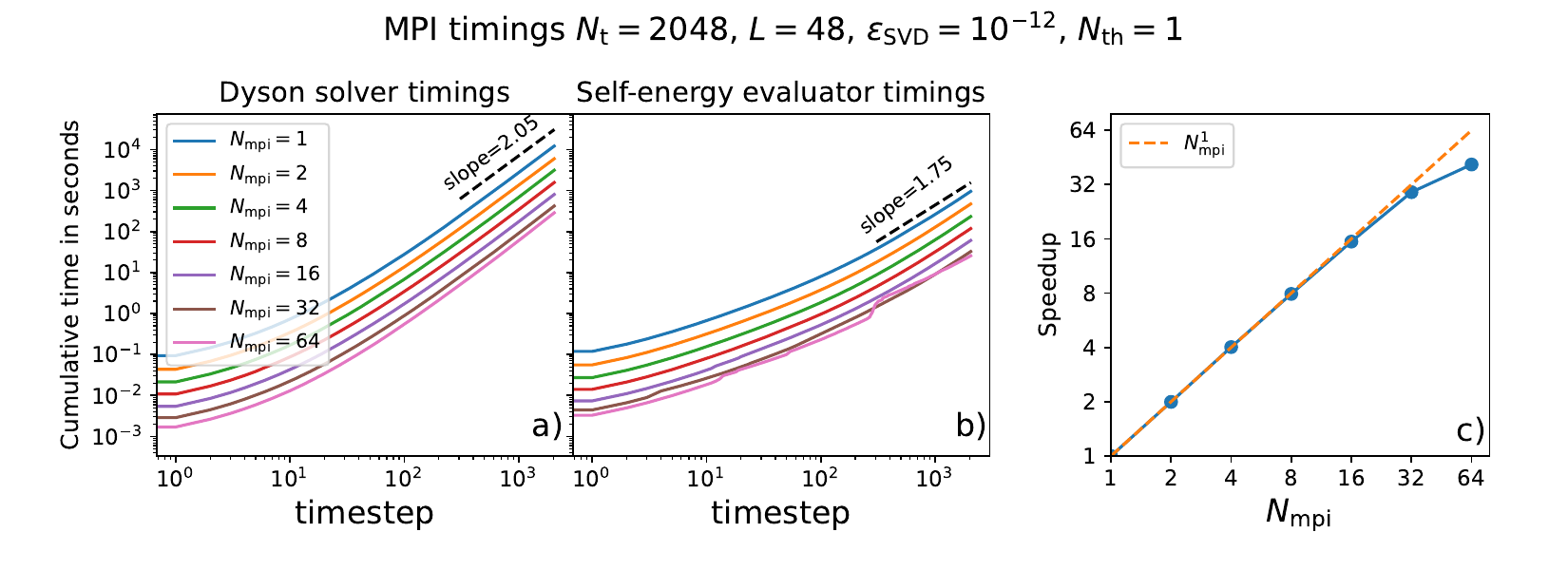}
	\includegraphics[width=0.85\textwidth]{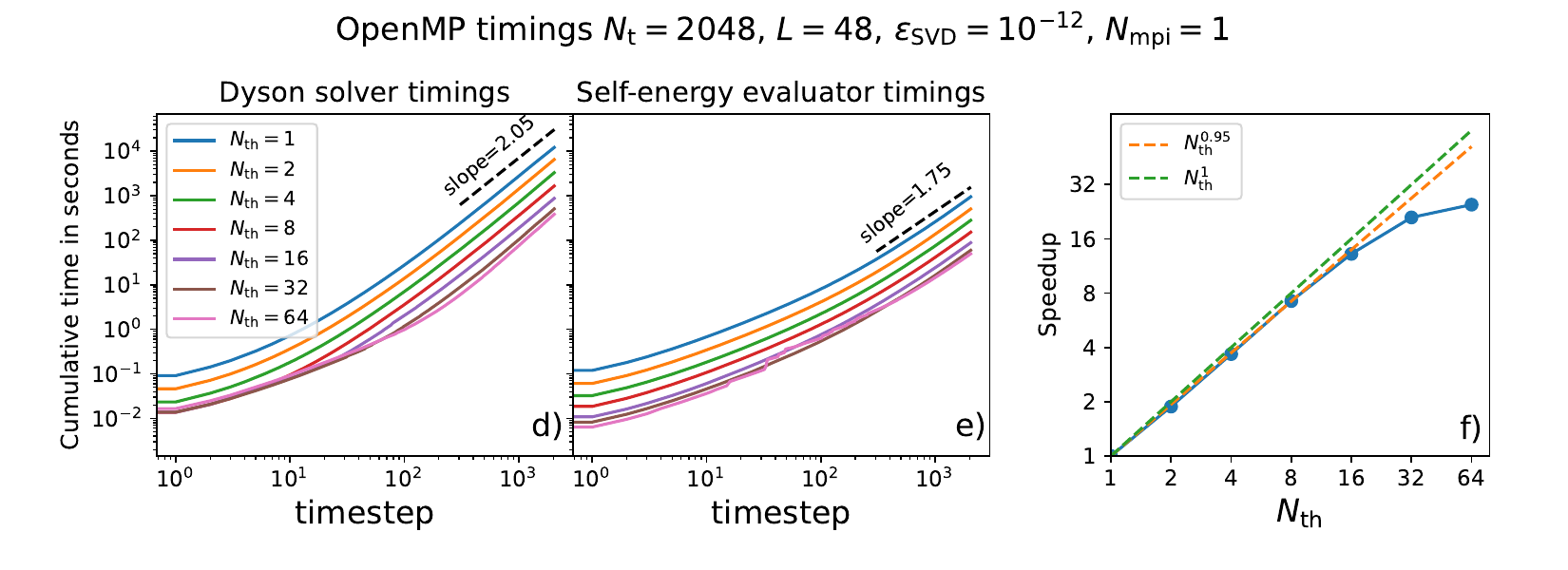}
	\centering
	\caption{Scaling of the Dyson solver (a,\,d) and the self-energy evaluator (b,\,e) for different numbers of MPI ranks and OpenMP threads, together with the relative speedup with the number of CPUs (c,\,f).  Upper panels: scaling with MPI ranks at $N_{\mathrm{th}}=1$.  Lower panels: scaling with OpenMP threads at $N_{\mathrm{mpi}}=1$.  The speedup is nearly linear in both cases.}
	\label{fig:small_example_scaling}
\end{figure*}

\subsection{Performance}
\label{sec:driven2d_performance}

\begin{figure*}[!h]
	\includegraphics[width=0.85\textwidth]{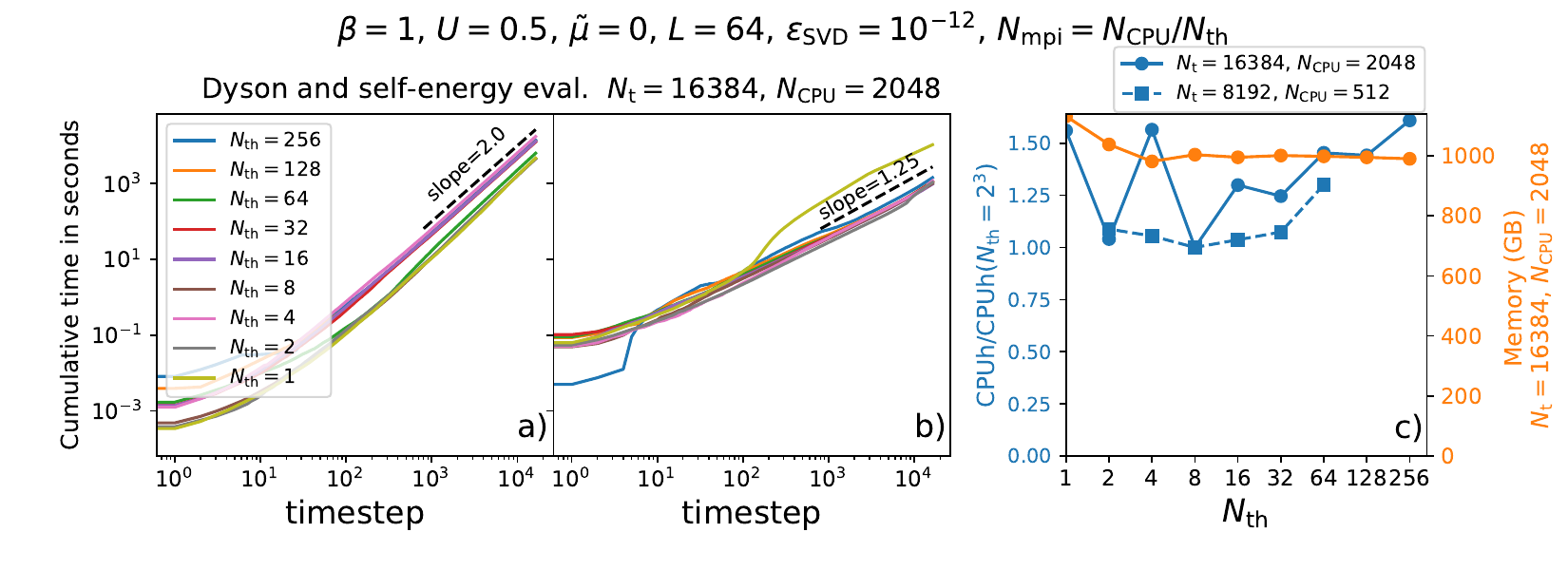}
	\centering
	\caption{Performance scaling of the hybrid MPI/OpenMP implementation for a fixed total number of CPUs ($N_{CPU} = 2048$). Cumulative computation time as a function of simulated timestep is shown for (a) the Dyson solver and (b) the self-energy evaluator across various OpenMP thread counts per MPI rank ($N_{th}$). (c) Computational metrics as a function of $N_{th}$. The left axis (blue) shows the total CPU hours normalized to the optimal case ($N_{th} = 8$) for two different simulation lengths. The right axis (orange) shows the total memory footprint for the $N_t = 16384$ calculation. For these hardware and calculation parameters, the optimal configuration that minimizes computational cost is $N_{th} = 8$.
	 }
	\label{fig:big_example_scaling}
\end{figure*}

In this section, we benchmark the hybrid MPI + OpenMP implementation.  To isolate the contributions of OpenMP and MPI, we first consider a smaller system with $L=48$ and $N_t=2048$, which can still be solved without parallelization. These results are shown in Fig.~\ref{fig:small_example_scaling}.  We then consider a large calculation with $L=64$ and $N_t=16384$, which requires the full hybrid implementation. These results are shown in Fig.~\ref{fig:big_example_scaling} and correspond to the calculations in Fig.~\ref{fig:ranks}.  In both cases, the integration scales as $N_t^{2.05}$, consistent with the expected scaling when the $\varepsilon$-ranks saturate at a maximal value, as confirmed in Fig.~\ref{fig:ranks}.  This behavior closely mirrors that of the single-site calculations in Sec.~\ref{sec:example_programs}.

The speedup plots in Fig.~\ref{fig:small_example_scaling} show the wall time required to reach a given timestep with the Dyson solver and the self-energy evaluator, as a function of CPU count, for an example where memory requirements do not necessitate parallelization.  The speedup relative to a single CPU is nearly perfectly linear up to 32 MPI processes on a single node.
The OpenMP scaling is slightly less efficient, but still reaches $N^{0.95}$ up to 16 threads before saturating.

Figure~\ref{fig:big_example_scaling} shows results for a system size and propagation time that require a large number of nodes and a hybrid parallelization strategy.
The efficiency depends sensitively on the total number of MPI ranks $N_{\mathrm{mpi}}$, the number of threads per rank $N_{\mathrm{th}}$, and their ratio.
At lower temperatures, for which the Green's functions decay slowly, a large number of time points ($N_\mathrm{t}>10000$) is needed for the current to fully decay.  Finite-size effects are also more prominent in this regime, requiring large lattice sizes ($N_k > 8000$), and memory constraints correspondingly demand more computational nodes.  For a fixed total number of CPUs, the ratio $N_{\mathrm{mpi}}/N_{\mathrm{th}}$ can be tuned to minimize computational cost.  Figure~\ref{fig:big_example_scaling}(c) shows the CPU hours and memory usage for different ratios at a fixed CPU count.  Memory grows slowly with $N_\mathrm{mpi}$ because each rank maintains its own copy of several auxiliary communication buffers (see Table~\ref{tab:comm_buffers}).  The scalings of the Dyson solver and self-energy evaluator are shown in Fig.~\ref{fig:big_example_scaling}(a) and (b), respectively.  The best performance is achieved with $N_{\mathrm{th}}=8$ threads per rank, a finding we have reproduced across two different examples and two computing clusters.

\section{Conclusions\label{sec:conclusions}}
We have presented \name{}, an open-source implementation of nonequilibrium Green's function propagation based on hierarchical low-rank compression of two-time objects. By combining high-order integration schemes with HODLR representations of the retarded and lesser components, the method reduces the cubic cost scaling and quadratic memory growth that traditionally limit KBE simulations. The discrete Lehmann representation provides an efficient treatment of imaginary-time quantities and thermal initial states. The resulting framework maintains controllable accuracy while substantially reducing computational cost, extending the accessible simulation times and system sizes for nonequilibrium correlated-electron problems.

We have demonstrated these capabilities in two representative applications: a driven superconducting system treated within dynamical mean-field theory, and the two-dimensional Hubbard model solved with the second Born approximation. In both cases, the compressed propagation exhibits favorable cost and memory requirements compared to conventional approaches while preserving the dynamical features of the underlying many-body problem. The modular design supports multiorbital systems, adaptive singular-value truncation, and hybrid MPI + OpenMP parallelization, making the code suitable for large-scale simulations on modern computing architectures.

The implementation provides a foundation for further extensions, including more advanced self-energy approximations, embedding schemes, and coupling to additional degrees of freedom. By reducing the computational barrier to two-time nonequilibrium simulations, \name{} opens the door to systematic studies of long-time dynamics, strong driving protocols, and complex multiorbital materials within a controlled many-body framework.

\section{Acknowledgments\label{sec:conclusions}}
T.B. was supported by the U.S. Department of Energy, Office of Science, Office of Advanced Scientific Computing Research, Department of Energy Computational Science Graduate Fellowship under Award No. DE-SC0020347 until Aug. 2023.  T.B. (after Aug 2023), and E.G. were supported by the U.S. Department of Energy, Office of Science, Office of Advanced Scientific Computing Research and Office of Basic Energy Sciences, Scientific Discovery through Advanced Computing (SciDAC) program under Award Number DE-SC0022088. As of June 1, 2025, E.G. was supported by the European Research Council, grant ERC-2023-AdG: 101142136 (Quantum Algorithms).  The Flatiron Institute is a division of the Simons Foundation. J. V. and J. Kov. acknowledge funding provided by the Institute of Physics Belgrade, through the grant by the Ministry of Science, Technological Development and Innovation of the Republic of Serbia. J. V. and J. Kov. acknowledge funding by the European Research Council, grant ERC-2022-StG: 101076100. D.G. is supported by the Slovenian Research and Innovation Agency (ARIS) under Programs No. P1-0044, No. J1-2455, and No. MN-0016-106. The authors gratefully acknowledge the HPC RIVR consortium [(www.hpc-rivr.si)](https://www.hpc-rivr.si) and EuroHPC JU [(eurohpc-ju.europa.eu)](https://eurohpc-ju.europa.eu/) for funding this research by providing computing resources of the HPC system Vega at the Institute of Information Science [(www.izum.si)](https://www.izum.si/en/home/). Computations were also performed on the PARADOX supercomputing facility (Scientific Computing Laboratory, Center for the Study of Complex Systems, Institute of Physics Belgrade).

\newpage
\onecolumn
\appendix

\section{Integral and derivative approximations}
\label{app:integration}
The numerical evaluation of derivatives, $\frac{d}{dt}y(t)|_T$, and integrals, $\int_0^T dt \, y(t)$, is split into two cases: $T\leq hk$ and $T>hk$.  In the former, we use $k^\mathrm{th}$-order polynomial differentiation and integration weights; in the latter, backward differentiation and Gregory integration~\cite{gregory_quadrature}.
To obtain the polynomial differentiation and integration weights, we begin with the polynomial interpolant of $y(t)$ that takes the values $y_j\equiv y(jh)$ on the equidistant grid: 
\begin{gather}
    \mathcal{P}[y_0,\dots,y_k](t)=\sum_{a,l=0}^k h^{-a}t^aP_{al}y_l\\
    P_{al}=(M^{-1})_{al} \text{ for 
    }M_{ja} = j^a
\end{gather}
for $0\leq j \leq k$.
The polynomial differentiation weights $D_{ml}$ are obtained by explicitly differentiating this interpolant:
\begin{gather}
\left.\frac{d y}{d t}\right|_{t=m h} \approx h^{-1} \sum_{l=0}^k D_{m l} y_l, \text { with } \\
D_{m l}=\sum_{a=1}^k P_{a l} a m^{a-1} .
\end{gather}
The polynomial integration weights $I_{mnl}$ are similarly obtained by explicitly integrating this interpolant:
\begin{gather}
 \int_{m h}^{n h} dt \, y(t) \approx h \sum_{l=0}^k I_{m nl} y_l, \text { with } \\
I_{m n l}=\sum_{a=0}^k P_{a l} \frac{n^{a+1}-m^{a+1}}{a+1}.
\end{gather}
These weights are then used for $T\leq hk$ to set up the so-called bootstrapping procedure, which is discussed in Sec.~\ref{ssec:dyson_class} and implemented in the functions shown in Table \ref{tab:Bootdyson}.

For the $T>hk$ case, we use the backward differentiation weights and Gregory weights for approximating the derivatives and integrals, respectively.  The backward differentiation weights $a_j$ are obtained from the polynomial differentiation weights via
\begin{gather}
\left.\frac{d y}{d t}\right|_{n h} \approx h^{-1} \sum_{j=0}^k a_j y_{n-j}\text{, where}\\
a_j=-D_{0j}.
\end{gather}
Lastly, the Gregory weights $\omega_i$ are best thought of as high-order edge corrections to the Riemann sum, where the corrections occur for $k+1$ points on each end of the sum
\begin{equation}
\int_0^{n h} d t \, y(t) \approx h \sum_{i=0}^k \omega_i y_i+h \sum_{i=k+1}^{n-k-1} y_i+h \sum_{i=0}^k \omega_i y_{n-i}.
\end{equation}
Since the integration weights differ from unity only near the endpoints, the evaluation of history integrals using the SVD representation is greatly simplified.  These weights are given by the sum
\begin{equation}
\omega_i=\sum_{j=0}^ib_{k+1-i}
\end{equation}
where $b_{i}$ are the weights defining the $k+1$ order Adams–Moulton method~\cite{gregory_quadrature}. These weights are then used for $T> hk$ for timestepping. The details of this procedure are described in Ref.~\cite{schuler2020}.

\section{Superconducting self-energy evaluation}
\label{app:selfen}
Here we provide an example self-energy and hybridization evaluation class for the DMFT superconducting problem of Section~\ref{sec:example_programs}.
The Fock diagram evaluation (Eq.~\ref{eq:SigmaF}) requires the density matrix, obtained from the Green's function.  In general the interaction $U$ may vary in time, and the implementation supports this case, although $U$ is constant in the example of Sec.~\ref{sec:example_programs}.
Because $U$ is a scalar and the \texttt{function} class stores matrix-valued functions, we access $U(t)$ as \texttt{U(tstp,0,0)}.
\begin{lstlisting}
void SC_gf2::solve_Sigma_Fock(int tstp, function &U, herm_matrix_hodlr &G, function &H) {
  ZMatrix rho(2,2);
  G.density_matrix(tstp, dlr_, rho);

  for(int i = 0; i < 2; i++) {
    H(tstp,i,1-i) = -U(tstp,0,0) * rho(i,1-i);
  }
}
\end{lstlisting}
Next, we implement the second-order diagrams of Eq.~\ref{eq:Sigma2B}.  When \texttt{tstp==-1}, only the Matsubara component is obtained; otherwise the retarded, lesser, and mixed components are computed.
\begin{lstlisting}
void SC_gf2::solve_Sigma(int tstp, function &U, herm_matrix_hodlr &G, herm_matrix_hodlr &Sigma) {
  if(tstp == -1) { Sigma_mat(U, G, Sigma); }
  else { Sigma_tstp(tstp, U, G, Sigma); }
}

void SC_gf2::Sigma_mat(function &U, herm_matrix_hodlr &G, herm_matrix_hodlr &Sigma) {
  DMatrix GM_reversed(G.ntau(), 4);
  G.get_mat_reversed(dlr_, GM_reversed);

  for(int tau = 0; tau < G.ntau(); tau++) {
    auto SM_map = Sigma.map_mat(tau);
    auto GM_map = G.map_mat(tau);
    for(int i = 0; i < 2; i++) {
      for(int j = 0; j < 2; j++) {
        // For reversed we index (i,j) using rowmajor ordering
        SM_map(i,j) += (U(-1,0,0) * U(-1,0,0) * GM_map(i,j) * GM_map(1-i,1-j) * GM_reversed(tau, (1-j)*2 + (1-i))).real();
        SM_map(i,j) -= (U(-1,0,0) * U(-1,0,0) * GM_map(i,1-j) * GM_map(1-i,j) * GM_reversed(tau, (1-j)*2 + (1-i))).real();
      }
    }
  }
}

void SC_gf2::Sigma_tstp(int tstp, function &U, herm_matrix_hodlr &G, herm_matrix_hodlr &Sigma) {
  for(int t = 0; t <= tstp; t++) {
    auto SR = Sigma.map_ret_curr(tstp, t);
    auto GR = G.map_ret_curr(tstp, t);
    auto SL = Sigma.map_les_curr(t, tstp);
    auto GL = G.map_les_curr(t, tstp);
    auto GG = GR - GL.adjoint();

    for(int i = 0; i < 2; i++) {
      for(int j = 0; j < 2; j++) {
        SL(i,j) += U(tstp,0,0) * U(t,0,0) * GL(i,j) * GL(1-i,1-j) * GG(1-j,1-i);
        SL(i,j) -= U(tstp,0,0) * U(t,0,0) * GL(i,1-j) * GG(1-j,1-i) * GL(1-i,j);
        // R(t,t') = >(t,t') - <(t,t') =  >(t,t') + <(t',t)^\dagger
        SR(i,j) += U(tstp,0,0) * U(t,0,0) * (GG(i,j) * GG(1-i,1-j) * GL(1-j,1-i) 
                                 + std::conj(GL(j,i) * GL(1-j,1-i) * GG(1-i,1-j)));
        SR(i,j) -= U(tstp,0,0) * U(t,0,0) * (GG(i,1-j) * GL(1-j,1-i) * GG(1-i,j) 
                                 + std::conj(GL(1-j,i) * GG(1-i,1-j) * GL(j,1-i)));
      }
    }
  }

  ZMatrix GVT(G.ntau(), 4);
  G.get_vt(tstp, dlr_, GVT);
  for(int tau = 0; tau < G.ntau(); tau++) {
    auto GTV_map = G.map_tv(tstp, tau);
    auto STV_map = Sigma.map_tv(tstp, tau);
    for(int i = 0; i < 2; i++) {
      for(int j = 0; j < 2; j++) {
        // For reversed we index (i,j) using rowmajor ordering
        STV_map(i,j) += U(tstp,0,0) * U(-1,0,0) * GTV_map(i,j) * GTV_map(1-i,1-j) * GVT(tau,(1-j)*2 + (1-i));
        STV_map(i,j) -= U(tstp,0,0) * U(-1,0,0) * GTV_map(i,1-j) * GVT(tau, (1-j)*2 + (1-i)) * GTV_map(1-i,j);
      }
    }
  }
}
\end{lstlisting}
Finally, we implement the hybridization function of Eq.~\ref{eq:Hybrid}. \texttt{sigma3\_} denotes the third Pauli matrix.
\begin{lstlisting}
void SC_gf2::solve_Delta(int tstp, function &t0, herm_matrix_hodlr &G, herm_matrix_hodlr &Sigma) {
  if(tstp == -1) { Delta_mat(t0, G, Sigma); }
  else { Delta_tstp(tstp, t0, G, Sigma); }
}

void SC_gf2::Delta_mat(function &t0, herm_matrix_hodlr &G, herm_matrix_hodlr &Sigma) {
  for(int tau = 0; tau < G.ntau(); tau++) {
    Sigma.map_mat(tau).noalias() += (t0.get_map(-1) * sigma3_ * G.map_mat(tau) * sigma3_ * t0.get_map(-1).conjugate()).real();
  }
}

void SC_gf2::Delta_tstp(int tstp, function &t0, herm_matrix_hodlr &G, herm_matrix_hodlr &Sigma) {
  ZMatrix sGs = ZMatrix::Zero(2,2);
  for(int t = 0; t <= tstp; t++) {
    sGs.noalias() = sigma3_ * G.map_ret_curr(tstp, t) * sigma3_;
    Sigma.map_ret_curr(tstp, t).noalias() += 0.5 * t0.get_map(tstp) * sGs * t0.get_map(t).conjugate();
    Sigma.map_ret_curr(tstp, t).noalias() += 0.5 * t0.get_map(tstp).conjugate() * sGs * t0.get_map(t);

    sGs.noalias() = sigma3_ * G.map_les_curr(t, tstp) * sigma3_;
    Sigma.map_les_curr(t, tstp).noalias() += 0.5 * t0.get_map(t) * sGs * t0.get_map(tstp).conjugate();
    Sigma.map_les_curr(t, tstp).noalias() += 0.5 * t0.get_map(t).conjugate() * sGs * t0.get_map(tstp);
  }

  for(int tau = 0; tau < G.ntau(); tau++) {
    sGs.noalias() = sigma3_ * G.map_tv(tstp, tau) * sigma3_;
    Sigma.map_tv(tstp, tau).noalias() += 0.5 * t0.get_map(tstp) * sGs * t0.get_map(-1).conjugate();
    Sigma.map_tv(tstp, tau).noalias() += 0.5 * t0.get_map(tstp).conjugate() * sGs * t0.get_map(-1);
  }
}
\end{lstlisting}

\section{Low-level \texttt{mpi\_comm} interface}
\label{app:mpi_comm_interface}
By default, \texttt{mpi\_comm} handles loading, unloading, and communication of Green's functions and self-energies automatically.  In cases where only a subset of orbital or Keldysh components needs to be communicated, a low-level interface allows users to implement custom protocols, as described below.

\begin{table}[h!]
    \centering
    \begin{tabular}{|c|c|}\hline
        \texttt{k\_rank\_map} & \makecell{Vector of size $N_{\mathrm{k}} $ that maps\\ a global k index to the MPI rank\\ responsible for that k-point.} \\\hline
        \texttt{my\_global\_k\_list} & \makecell{Vector of size $N^{\mathrm{loc}}_{\mathrm{k}}$ that maps\\ local k-point indices at each  MPI\\ rank to the global k-point indices.} \\\hline
        \texttt{Nk\_per\_rank} & \makecell{Vector of size $N_{\mathrm{mpi}}$ that contains\\ the number of k points\\ handled by each MPI rank} \\\hline
        \texttt{init\_tau\_per\_rank} & \makecell{Vector of size $N_{\mathrm{mpi}}$ that contains\\the initial $\tau$ index at each rank.} \\\hline
		\texttt{Ntau\_per\_rank} & \makecell{Vector of size $N_{\mathrm{mpi}}$ that contains \\the number of $\tau$ points \\ handled by each MPI rank} \\\hline
    \end{tabular}
    \caption{Vectors inside \texttt{mpi\_comm} describing the workload distribution among MPI ranks.}
    \label{tab:rank_distribution}
\end{table}

\begin{table}[h!]
    \centering
    \begin{tabular}{|c|c|}\hline
        \texttt{tau\_alltp\_buff} & \makecell{Buffer that contains the data at all\\ $\tau$ points and a set of $\mathbf{k}$-points\\ local to the MPI rank.} \\\hline
        \texttt{t\_alltp\_buff} & \makecell{Buffer that contains the data at all\\ $t'$ points and a set of $\mathbf{k}$-points\\ local to the MPI rank.} \\\hline
        \texttt{tau\_allk\_buff} & \makecell{Buffer that contains the data at all\\ $\mathbf{k}$-points and a set of $\tau$ points\\ local to the MPI rank.} \\\hline
        \texttt{t\_allk\_buff} & \makecell{Buffer that contains the data at all\\ $\mathbf{k}$-points and a set of $t'$ points\\ local to the MPI rank.} \\\hline
    \end{tabular}
    \caption{Intermediate communication buffers inside \texttt{mpi\_comm}.}
    \label{tab:comm_buffers}
\end{table}

\begin{table}[h!]
    \centering
    \begin{tabular}{|c|c|}\hline
        \texttt{alltp\_buff\_index} & \makecell{Computes the index \\in the {\tt alltp\_buff} buffers.} \\\hline
        \texttt{t\_allk\_buff\_index} & \makecell{Computes the index \\in the {\tt t\_allk\_buff} buffer.} \\\hline
        \texttt{tau\_allk\_buff\_index} & \makecell{Computes the index\\ in the {\tt tau\_allk\_buff} buffer.} \\\hline
    \end{tabular}
    \caption{Helper indexing functions inside \texttt{mpi\_comm}.}
    \label{tab:mpi_helpers}
\end{table}

The workload distribution and communication details are managed by the \texttt{mpi\_comm} class, whose main components are summarized in Tables~\ref{tab:rank_distribution}--\ref{tab:mpi_helpers}. At the start of the program, each MPI rank initializes an instance with
\begin{lstlisting}
mpi_comm comm(global_Nk, Nt, r, nao, max_component_size);
\end{lstlisting}
where \texttt{Nk}~$=N_{\mathrm{k}}$ is the total number of $\mathbf{k}$-points, \texttt{Nt}~$=N_{\mathrm{t}}$ is the total number of time points, \texttt{r} is the number of DLR imaginary-time nodes used by the \texttt{dyson} class, \texttt{nao} is the number of orbitals, and \texttt{max\_component\_size} is the maximum number of Keldysh and/or orbital components sent per communication (default~2). During initialization, the communicator assigns blocks of $\mathbf{k}$-points and $\tau$~points to each rank (Table~\ref{tab:rank_distribution}), whereas real-time $t'$~points are distributed on the fly at each timestep~$t$, since the number of $\tau$~points is fixed but the number of $t'$~points grows with each timestep. Several buffers are also allocated for communication and self-energy evaluation (Table~\ref{tab:comm_buffers}). In the self-consistent loop, the Green's functions are extracted from the \texttt{herm\_matrix\_hodlr} objects into \texttt{alltp\_buff} and communicated to \texttt{allk\_buff} in the downward direction shown in Fig.~\ref{fig:communication}; both steps are handled by \texttt{mpi\_get\_and\_comm}. The self-energy is then evaluated inside \texttt{allk\_buff} and communicated back to \texttt{alltp\_buff} in the upward direction (Fig.~\ref{fig:communication}), after which the results are written into the \texttt{herm\_matrix\_hodlr} objects; these steps are handled by \texttt{mpi\_comm\_and\_set}. Because \texttt{MPI\_Alltoallv} requires one-dimensional arrays, the indexing into the communication buffers is nontrivial; the helper functions in Table~\ref{tab:mpi_helpers} perform the necessary index calculations.

The MPI communicator is agnostic to the internal layout of the Green's functions and self-energies. The user therefore supplies lambda expressions specifying how to read and write each component; these lambdas are collected into vectors and passed to the communicator methods, which iterate over them to perform the extraction or writing. The user may define any number of lambdas to access any Keldysh component or orbital of interest. The only constraint is that \texttt{max\_component\_size} must be set to at least the largest number of components sent in a single communication. Therefore, if the user wants to send $n$ Keldysh components of the Green's function in one communication, {\tt max\_component\_size} must be at least $n$. If fewer components are sent in a subsequent call, the existing buffer space suffices; if more are sent than \texttt{max\_component\_size}, an error is thrown.

For example, computing the lesser self-energy from Eq.~\ref{eq:BornApproxLG} requires both $G^>$ and $G^<$. The corresponding get lambdas are
\begin{lstlisting}
//returns G^<(t,t') at local k index and real times t,t'
auto getLess = [&corrK_rank, &size]
(int local_ki, int ti, int tpi)
{
  hodlr::ZMatrix tmp(size,size);
  corrK_rank[local_ki]->G_.get_les(ti,tpi,tmp);
  return tmp(0,0);
};

//returns G^>(t,t') at local k index and real times t,t'
auto getGreat = [&corrK_rank, &size](int local_ki, int ti, int tpi){
  hodlr::ZMatrix tmpRet(size,size),tmpLess(size,size),tmpGreat(size,size);

  corrK_rank[local_ki]->G_.get_les(ti,tpi,tmpLess);
  corrK_rank[local_ki]->G_.get_ret(ti,tpi,tmpRet);
  tmpGreat(0,0) = tmpRet(0,0)+tmpLess(0,0);

  return tmpGreat(0,0);
};

//vector of lambdas for Lesser and Greater components
std::vector<std::function<std::complex<double>(int, int, int)>> getsLG = {getLess,getGreat};
\end{lstlisting}
After computing and communicating the self-energy, the communicator needs to write the results back into the self-energy objects. The following set lambdas write $\Sigma^{\mathrm{R}}_{\mathbf{k}}(t,t')$ and $\Sigma^{<}_{\mathbf{k}}(t,t')$:
\begin{lstlisting}
//sets Sigma^<(t,t') at local k index and real time t for all t'<=t
auto setLess = [&corrK_rank, &size](int k, int ti, std::vector<std::complex<double>> &Sigma){
  std::vector<std::complex<double>> Sigmac(ti+1);
  for(int j = 0; j<=ti; j++){
  	Sigmac[j] = -std::conj(Sigma[j]);
  }
  hodlr::ZMatrixMap(corrK_rank[k]->Sigma_.curr_timestep_les_ptr(0,ti), (ti + 1) * size, size).noalias() = hodlr::ZMatrixMap(Sigmac.data(), (ti + 1) * size, size);
};

//sets Sigma^R(t,t') at local k index and real time t for all t'>=t
auto setRet = [&corrK_rank, &size](int k, int ti, std::vector<std::complex<double>> &Sigma)
{
  hodlr::ZMatrixMap(corrK_rank[k]->Sigma_.curr_timestep_ret_ptr(ti,0), (ti + 1) * size, size).noalias() = hodlr::ZMatrixMap(Sigma.data(), (ti + 1) * size, size);
};

std::vector<std::function<void(int, int, std::vector<std::complex<double>>&)>> setsLR = {setLess,setRet};
\end{lstlisting} 	
The vectors \texttt{getsLG} and \texttt{setsLR} are passed to the communicator methods, which iterate over their elements and perform the extraction or writing in the appropriate communication direction (Fig.~\ref{fig:communication}). In this example, \texttt{max\_component\_size} must be at least~2, since two components of each quantity are communicated.

For general complex time indices $t$ and $t'$ and Keldysh component index $\eta$, the get and set lambdas must satisfy the following requirements:
\begin{itemize}
	\item A get lambda returns $G^\eta_{\mathbf{k}}(t,t')$ at a local $\mathbf{k}$ index, with signature
	\texttt{std::complex<double>(int local\_ki, int ti, int tpi)}.
	
	\item A set lambda writes $\Sigma^\eta_{\mathbf{k}}(t,t')$ at a local $\mathbf{k}$ index for an array of $t'\leq t$ points, with signature
	\texttt{void(int local\_ki, int ti, std::vector<std::complex<double>> \&Sigma\_tp)}.

	\item Both get and set lambdas must use the timestep $t$ as their first time index, including for the Matsubara case where $t=-1$.
\end{itemize}

\newpage

\bibliographystyle{unsrt}
\bibliography{sample}

@article{schuler2020,
	author = {Sch{\"u}ler, Michael and Gole{\v{z}}, Denis and Murakami, Yuta and Bittner, Nikolaj and Herrmann, Andreas and Strand, Hugo UR and Werner, Philipp and Eckstein, Martin},
	journal = {Comput. Phys. Commun.},
	pages = {107484},
	publisher = {Elsevier},
	title = {NESSi: The Non-Equilibrium Systems Simulation package},
	volume = {257},
	year = {2020}}

@misc{sroda2024,
      title={High-resolution nonequilibrium $GW$ calculations based on quantics tensor trains}, 
      author={Maksymilian Środa and Ken Inayoshi and Hiroshi Shinaoka and Philipp Werner},
      year={2024},
      eprint={2412.14032},
      archivePrefix={arXiv},
      primaryClass={cond-mat.str-el},
      url={https://arxiv.org/abs/2412.14032}, 
}

@inbook{ballani16,
	author = {Ballani, Jonas and Kressner, Daniel},
	booktitle = {Exploiting Hidden Structure in Matrix Computations: Algorithms and Applications},
	editor = {Benzi, Michele and Simoncini, Valeria},
	pages = {161--209},
	publisher = {Springer},
	title = {Matrices with Hierarchical Low-Rank Structures},
	year = {2016}}

@article{basov2017towards,
	author = {Basov, DN and Averitt, RD and Hsieh, D},
	journal = {Nat. Mater.},
	number = {11},
	pages = {1077--1088},
	publisher = {Nature Publishing Group},
	title = {Towards properties on demand in quantum materials},
	volume = {16},
	year = {2017}}

@article{giannetti2016ultrafast,
	author = {Giannetti, Claudio and Capone, Massimo and Fausti, Daniele and Fabrizio, Michele and Parmigiani, Fulvio and Mihailovic, Dragan},
	journal = {Adv. Phys.},
	number = {2},
	pages = {58--238},
	publisher = {Taylor \& Francis},
	title = {Ultrafast optical spectroscopy of strongly correlated materials and high-temperature superconductors: a non-equilibrium approach},
	volume = {65},
	year = {2016}}

@article{ligges2018ultrafast,
	author = {Ligges, Manuel and Avigo, Isabella and Gole{\v{z}}, Denis and Strand, HUR and Beyazit, Y and Hanff, K and Diekmann, F and Stojchevska, L and Kall{\"a}ne, M and Zhou, P and others},
	journal = {Phys. Rev. Lett.},
	number = {16},
	pages = {166401},
	publisher = {APS},
	title = {Ultrafast Doublon Dynamics in Photoexcited 1 T-TaS 2},
	volume = {120},
	year = {2018}}

@book{stefanucci2013nonequilibrium,
	author = {Stefanucci, Gianluca and Van Leeuwen, Robert},
	publisher = {Cambridge University Press},
	title = {Nonequilibrium many-body theory of quantum systems: a modern introduction},
	year = {2013}}

@book{haug2008quantum,
	author = {Haug, Hartmut and Jauho, Antti-Pekka},
	edition = {2},
	publisher = {Springer},
	title = {Quantum kinetics in transport and optics of semiconductors},
	year = {2008}}

@book{marques2012fundamentals,
	author = {Marques, Miguel AL and Maitra, Neepa T and Nogueira, Fernando MS and Gross, Eberhard KU and Rubio, Angel},
	publisher = {Springer},
	title = {Fundamentals of time-dependent density functional theory},
	year = {2012}}

@article{schollwock2011density,
	author = {Schollw{\"o}ck, Ulrich},
	journal = {Ann. Phys.},
	number = {1},
	pages = {96--192},
	publisher = {Elsevier},
	title = {The density-matrix renormalization group in the age of matrix product states},
	volume = {326},
	year = {2011}}

@article{werner2010,
	author = {Werner, Philipp and Oka, Takashi and Eckstein, Martin and Millis, Andrew J.},
	issue = {3},
	journal = {Phys. Rev. B},
	month = {Jan},
	numpages = {11},
	pages = {035108},
	publisher = {American Physical Society},
	title = {Weak-coupling quantum Monte Carlo calculations on the Keldysh contour: Theory and application to the current-voltage characteristics of the Anderson model},
	volume = {81},
	year = {2010}}

@article{cohen2015,
	author = {Cohen, Guy and Gull, Emanuel and Reichman, David R. and Millis, Andrew J.},
	issue = {26},
	journal = {Phys. Rev. Lett.},
	month = {Dec},
	numpages = {5},
	pages = {266802},
	publisher = {American Physical Society},
	title = {Taming the Dynamical Sign Problem in Real-Time Evolution of Quantum Many-Body Problems},
	volume = {115},
	year = {2015}}

@article{gull2011,
	author = {Gull, Emanuel and Reichman, David R. and Millis, Andrew J.},
	issue = {8},
	journal = {Phys. Rev. B},
	month = {Aug},
	numpages = {4},
	pages = {085134},
	publisher = {American Physical Society},
	title = {Numerically exact long-time behavior of nonequilibrium quantum impurity models},
	volume = {84},
	year = {2011}}

@article{aoki2014,
	author = {Aoki, Hideo and Tsuji, Naoto and Eckstein, Martin and Kollar, Marcus and Oka, Takashi and Werner, Philipp},
	journal = {Rev. Mod. Phys.},
	number = {2},
	pages = {779},
	publisher = {APS},
	title = {Nonequilibrium dynamical mean-field theory and its applications},
	volume = {86},
	year = {2014}}

@article{fausti2011,
	author = {Fausti, Daniele and Tobey, RI and Dean, Nicky and Kaiser, Stefan and Dienst, A and Hoffmann, Matthias C and Pyon, S and Takayama, T and Takagi, H and Cavalleri, Andrea},
	journal = {Science},
	number = {6014},
	pages = {189--191},
	publisher = {American Association for the Advancement of Science},
	title = {Light-induced superconductivity in a stripe-ordered cuprate},
	volume = {331},
	year = {2011}}

@article{mitrano2016,
	author = {Mitrano, Matteo and Cantaluppi, Alice and Nicoletti, Daniele and Kaiser, Stefan and Perucchi, A and Lupi, S and Di Pietro, P and Pontiroli, D and Ricc{\`o}, M and Clark, Stephen R and others},
	journal = {Nature},
	number = {7591},
	pages = {461--464},
	publisher = {Nature Publishing Group},
	title = {Possible light-induced superconductivity in K 3 C 60 at high temperature},
	volume = {530},
	year = {2016}}

@article{schuler2018,
	author = {Sch\"uler, Michael and Eckstein, Martin and Werner, Philipp},
	issue = {24},
	journal = {Phys. Rev. B},
	month = {Jun},
	numpages = {14},
	pages = {245129},
	publisher = {American Physical Society},
	title = {Truncating the memory time in nonequilibrium dynamical mean field theory calculations},
	volume = {97},
	year = {2018}}

@article{lipavsky1986,
	author = {Lipavsk{\`y}, P and {\v{S}}pi{\v{c}}ka, V and Velick{\`y}, B},
	journal = {Phys. Rev. B},
	number = {10},
	pages = {6933},
	publisher = {APS},
	title = {Generalized Kadanoff-Baym ansatz for deriving quantum transport equations},
	volume = {34},
	year = {1986}}

@article{kalvova2019,
	author = {Kalvov{\'a}, And{\v{e}}la and Velick{\`y}, Bed{\v{r}}ich and {\v{S}}pi{\v{c}}ka, V{\'a}clav},
	journal = {Phys. Status Solidi B},
	number = {7},
	pages = {1800594},
	publisher = {Wiley Online Library},
	title = {Beyond the Generalized Kadanoff--Baym Ansatz},
	volume = {256},
	year = {2019}}

@article{tuovinen2020comparing,
	author = {Tuovinen, Riku and Gole{\v{z}}, Denis and Eckstein, Martin and Sentef, Michael A},
	journal = {Phys. Rev. B},
	number = {11},
	pages = {115157},
	publisher = {APS},
	title = {Comparing the generalized Kadanoff-Baym ansatz with the full Kadanoff-Baym equations for an excitonic insulator out of equilibrium},
	volume = {102},
	year = {2020}}

@book{kadanoff1962quantum,
	author = {Kadanoff, Leo P and Baym, Gordon A},
	publisher = {W.A. Benjamin},
	title = {Quantum Statistical Mechanics},
	year = {1962}}

@book{balzer2012nonequilibrium,
	author = {Balzer, Karsten and Bonitz, Michael},
	publisher = {Springer},
	title = {Nonequilibrium Green's Functions Approach to Inhomogeneous Systems},
	year = {2012}}

@phdthesis{balzer2011solving,
  title={Solving the two-time Kadanoff-Baym equations: Application to model atoms and molecules},
  author={Balzer, Karsten},
  year={2011}
}

@article{bassi2024,
  title={Learning nonlinear integral operators via recurrent neural networks and its application in solving integro-differential equations},
  author={Bassi, Hardeep and Zhu, Yuanran and Liang, Senwei and Yin, Jia and Reeves, Cian C and Vl{\v{c}}ek, Vojt{\v{e}}ch and Yang, Chao},
  journal={Machine Learning with Applications},
  volume={15},
  pages={100524},
  year={2024},
  publisher={Elsevier}
}

@article{georges1996dynamical,
	author = {Georges, Antoine and Kotliar, Gabriel and Krauth, Werner and Rozenberg, Marcelo J},
	journal = {Rev. Mod. Phys.},
	number = {1},
	pages = {13},
	publisher = {APS},
	title = {Dynamical mean-field theory of strongly correlated fermion systems and the limit of infinite dimensions},
	volume = {68},
	year = {1996}}

@article{li2019long,
	author = {Li, Jiajun and Gole{\v{z}}, Denis and Werner, Philipp and Eckstein, Martin},
	journal = {arXiv preprint arXiv:1908.08693},
	title = {Long-range $\eta$-pairing in photodoped Mott insulators},
	year = {2019}}

@article{perfetto2018ultrafast,
	author = {Perfetto, E and Sangalli, D and Marini, A and Stefanucci, G},
	journal = {J. Phys. Chem. Lett.},
	number = {6},
	pages = {1353--1358},
	publisher = {ACS Publications},
	title = {Ultrafast charge migration in xuv photoexcited phenylalanine: A first-principles study based on real-time nonequilibrium green's functions},
	volume = {9},
	year = {2018}}

@article{molina2017ab,
	author = {Molina-S{\'a}nchez, Alejandro and Sangalli, Davide and Wirtz, Ludger and Marini, Andrea},
	journal = {Nano Lett.},
	number = {8},
	pages = {4549--4555},
	publisher = {ACS Publications},
	title = {Ab initio calculations of ultrashort carrier dynamics in two-dimensional materials: valley depolarization in single-layer wse2},
	volume = {17},
	year = {2017}}

@article{sangalli2019many,
	author = {Sangalli, D and Ferretti, A and Miranda, H and Attaccalite, Claudio and Marri, I and Cannuccia, Elena and Melo, P and Marsili, M and Paleari, F and Marrazzo, A and others},
	journal = {J. Phys. Condens. Matter},
	number = {32},
	pages = {325902},
	publisher = {IOP Publishing},
	title = {Many-body perturbation theory calculations using the yambo code},
	volume = {31},
	year = {2019}}

@article{marini2009yambo,
	author = {Marini, Andrea and Hogan, Conor and Gr{\"u}ning, Myrta and Varsano, Daniele},
	journal = {Comput. Phys. Commun.},
	number = {8},
	pages = {1392--1403},
	publisher = {Elsevier},
	title = {Yambo: an ab initio tool for excited state calculations},
	volume = {180},
	year = {2009}}

@article{konstantinov1961diagram,
  title={A Diagram Technique for Evaluating Transport Quantities},
  author={Konstantinov, O. V. and Perel', V. I.},
  journal={Sov. Phys. JETP},
  volume={12},
  number={1},
  pages={142--149},
  year={1961},
  month={January}
}

@article{Dong_Krivenko_Kleinhenz_Antipov_Cohen_Gull_2017, title={Quantum Monte Carlo solution of the dynamical mean field equations in real time}, volume={96}, DOI={10.1103/PhysRevB.96.155126}, abstractNote={We present real-time inchworm quantum Monte Carlo results for single-site dynamical mean field theory on an infinite coordination number Bethe lattice. Our numerically exact results are obtained on the L-shaped Keldysh contour and, being evaluated in real time, avoid the analytic continuation issues typically encountered in Monte Carlo calculations. Our results show that inchworm Monte Carlo methods have now reached a state where they can be used as dynamical mean field impurity solvers and the dynamical sign problem can be overcome. As nonequilibrium problems can be simulated at the same cost, we envisage the main use of these methods as dynamical mean field solvers for time-dependent problems far from equilibrium.}, number={15}, journal={Physical Review B}, publisher={American Physical Society}, author={Dong, Qiaoyuan and Krivenko, Igor and Kleinhenz, Joseph and Antipov, Andrey E. and Cohen, Guy and Gull, Emanuel}, year={2017}, month=oct, pages={155126} }

@article{Cohen_Gull_Reichman_Millis_Rabani_2013, title={Numerically exact long-time magnetization dynamics at the nonequilibrium Kondo crossover of the Anderson impurity model}, volume={87}, DOI={10.1103/PhysRevB.87.195108}, abstractNote={We investigate the dynamical and steady-state spin response of the nonequilibrium Anderson model to magnetic fields, bias voltage, and temperature using a numerically exact method combining a bold-line quantum Monte Carlo technique with a memory function formalism. We obtain converged results in a range of previously inaccessible regimes, in particular calculating the spin dynamics for a range of temperatures down to the crossover to the Kondo domain. We provide predictions for nonequilibrium phenomena, including nonmonotonic temperature dependence of observables at high bias voltage and oscillatory quench dynamics at high magnetic fields.}, number={19}, journal={Physical Review B}, publisher={American Physical Society}, author={Cohen, Guy and Gull, Emanuel and Reichman, David R. and Millis, Andrew J. and Rabani, Eran}, year={2013}, month=may, pages={195108} }

@article{Cohen_Rabani_2011, title={Memory effects in nonequilibrium quantum impurity models}, volume={84}, DOI={10.1103/PhysRevB.84.075150}, abstractNote={Memory effects play a key role in the dynamics of strongly correlated systems driven out of equilibrium. In this paper, we explore the nature of memory in the nonequilibrium Anderson impurity model. The Nakajima-Zwanzig-Mori formalism is used to derive an exact generalized quantum master equation for the reduced density matrix of the interacting quantum dot, which includes a non-Markovian memory kernel. A real-time path integral formulation is developed in which all diagrams are stochastically sampled in order to numerically evaluate the memory kernel. We explore the effects of temperature down to the Kondo regime, as well as the role of source-drain-bias voltage and bandwidth on the memory. Typically, the memory decays on time scales significantly shorter than the dynamics of the reduced density matrix itself, yet under certain conditions, it develops a low magnitude but long-ranged tail. In addition, we address the conditions required for the existence, uniqueness, and stability of a steady state.}, number={7}, journal={Physical Review B}, publisher={American Physical Society}, author={Cohen, Guy and Rabani, Eran}, year={2011}, month=aug, pages={075150} }

@article{Erpenbeck_Zhu_Yu_Zhang_Gerum_Goulko_Yang_Cohen_Gull_2025, title={Compact representation and long-time extrapolation of real-time data for quantum systems}, url={http://arxiv.org/abs/2506.13760}, DOI={10.48550/arXiv.2506.13760}, abstractNote={Representing real-time data as a sum of complex exponentials provides a compact form that enables both denoising and extrapolation. As a fully data-driven method, the Estimation of Signal Parameters via Rotational Invariance Techniques (ESPRIT) algorithm is agnostic to the underlying physical equations, making it broadly applicable to various observables and experimental or numerical setups. In this work, we consider applications of the ESPRIT algorithm primarily to extend real-time dynamical data from simulations of quantum systems. We evaluate ESPRIT’s performance in the presence of noise and compare it to other extrapolation methods. We demonstrate its ability to extract information from short-time dynamics to reliably predict long-time behavior and determine the minimum time interval required for accurate results. We discuss how this insight can be leveraged in numerical methods that propagate quantum systems in time, and show how ESPRIT can predict infinite-time values of dynamical observables, offering a purely data-driven approach to characterizing quantum phases.}, note={arXiv:2506.13760 [cond-mat]}, number={arXiv:2506.13760}, publisher={arXiv}, author={Erpenbeck, Andre and Zhu, Yuanran and Yu, Yang and Zhang, Lei and Gerum, Richard and Goulko, Olga and Yang, Chao and Cohen, Guy and Gull, Emanuel}, year={2025}, month=june }

@article{Erpenbeck_Blommel_Zhang_Lin_Cohen_Gull_2024, title={Steady-state properties of multi-orbital systems using quantum Monte Carlo}, volume={161}, ISSN={0021-9606}, DOI={10.1063/5.0226253}, abstractNote={A precise dynamical characterization of quantum impurity models with multiple interacting orbitals is challenging. In quantum Monte Carlo methods, this is embodied by sign problems. A dynamical sign problem makes it exponentially difficult to simulate long times. A multi-orbital sign problem generally results in a prohibitive computational cost for systems with multiple impurity degrees of freedom even in static equilibrium calculations. Here, we present a numerically exact inchworm method that simultaneously alleviates both sign problems, enabling simulation of multi-orbital systems directly in the equilibrium or nonequilibrium steady-state. The method combines ideas from the recently developed steady-state inchworm Monte Carlo framework [Erpenbeck et al., Phys. Rev. Lett. 130, 186301 (2023)] with other ideas from the equilibrium multi-orbital inchworm algorithm [Eidelstein et al., Phys. Rev. Lett. 124, 206405 (2020)]. We verify our method by comparison with analytical limits and numerical results from previous methods.}, number={9}, journal={The Journal of Chemical Physics}, author={Erpenbeck, A. and Blommel, T. and Zhang, L. and Lin, W.-T. and Cohen, G. and Gull, E.}, year={2024}, month=sept, pages={094104} }

@article{Dong_Gull_Strand_2022, title={Excitations and spectra from equilibrium real-time Green’s functions}, volume={106}, DOI={10.1103/PhysRevB.106.125153}, abstractNote={The real-time contour formalism for Green’s functions provides time-dependent information of quantum many-body systems. In practice, the long-time simulation of systems with a wide range of energy scales is challenging due to both the storage requirements of the discretized Green’s function and the computational cost of solving the Dyson equation. In this paper, we apply a real-time discretization based on a piecewise high-order orthogonal-polynomial expansion to address these issues. We present a superconvergent algorithm for solving the real-time equilibrium Dyson equation using the Legendre spectral method and the recursive algorithm for Legendre convolution. We show that the compact high-order discretization in combination with our Dyson solver enables long-time simulations using far fewer discretization points than needed in conventional multistep methods. As a proof of concept, we compute the molecular spectral functions of H2, LiH, He2, and C6⁢H4⁡O2 using self-consistent second-order perturbation theory and compare the results with standard quantum chemistry methods as well as the auxiliary second-order Green’s function perturbation theory method.}, number={12}, journal={Physical Review B}, publisher={American Physical Society}, author={Dong, Xinyang and Gull, Emanuel and Strand, Hugo U. R.}, year={2022}, month=sept, pages={125153} }

@article{Erpenbeck_Gull_Cohen_2023, title={Quantum Monte Carlo Method in the Steady State}, volume={130}, DOI={10.1103/PhysRevLett.130.186301}, abstractNote={We present a numerically exact steady-state inchworm Monte Carlo method for nonequilibrium quantum impurity models. Rather than propagating an initial state to long times, the method is directly formulated in the steady state. This eliminates any need to traverse the transient dynamics and grants access to a much larger range of parameter regimes at vastly reduced computational costs. We benchmark the method on equilibrium Green’s functions of quantum dots in the noninteracting limit and in the unitary limit of the Kondo regime. We then consider correlated materials described with dynamical mean field theory and driven away from equilibrium by a bias voltage. We show that the response of a correlated material to a bias voltage differs qualitatively from the splitting of the Kondo resonance observed in bias-driven quantum dots.}, number={18}, journal={Physical Review Letters}, publisher={American Physical Society}, author={Erpenbeck, A. and Gull, E. and Cohen, G.}, year={2023}, month=may, pages={186301} }

@article{perfetto2018cheers,
	author = {Perfetto, Enrico and Stefanucci, Gianluca},
	journal = {J. Phys. Condens. Matter},
	number = {46},
	pages = {465901},
	publisher = {IOP Publishing},
	title = {CHEERS: a tool for correlated hole-electron evolution from real-time simulations},
	volume = {30},
	year = {2018}}

@article{zhou2019nonequilibrium,
	author = {Zhou, Faran and Williams, Joseph and Malliakas, Christos D and Kanatzidis, Mercouri G and Kemper, Alexander F and Ruan, Chong-Yu},
	journal = {arXiv preprint arXiv:1904.07120},
	title = {Nonequilibrium dynamics of spontaneous symmetry breaking into a hidden state of charge-density wave},
	year = {2019}}

@article{blommel2024,
  title = {Adaptive time stepping for the two-time integro-differential Kadanoff-Baym equations},
  author = {Blommel, Thomas and Gardner, David J. and Woodward, Carol S. and Gull, Emanuel},
  journal = {Phys. Rev. B},
  volume = {110},
  issue = {20},
  pages = {205134},
  numpages = {13},
  year = {2024},
  month = {Nov},
  publisher = {American Physical Society},
  doi = {10.1103/PhysRevB.110.205134},
  url = {https://link.aps.org/doi/10.1103/PhysRevB.110.205134}
}

@article{gregory_quadrature,
    author = {WOLKENFELT, P. H. M.},
    title = {The Construction of Reducible Quadrature Rules for Volterra Integral and Integro-differential Equations},
    journal = {IMA Journal of Numerical Analysis},
    volume = {2},
    number = {2},
    pages = {131-152},
    year = {1982},
    month = {04},
    issn = {0272-4979},
    doi = {10.1093/imanum/2.2.131},
    url = {https://doi.org/10.1093/imanum/2.2.131},
    eprint = {https://academic.oup.com/imajna/article-pdf/2/2/131/2267756/2-2-131.pdf},
}

@article{KAYE_libdlr,
title = {libdlr: Efficient imaginary time calculations using the discrete {L}ehmann representation},
_journal = {Computer Physics Communications},
journal = {Comput. Phys. Commun.},
volume = {280},
pages = {108458},
year = {2022},
issn = {0010-4655},
doi = {https://doi.org/10.1016/j.cpc.2022.108458},
url = {https://www.sciencedirect.com/science/article/pii/S0010465522001771},
author = {Jason Kaye and Kun Chen and Hugo U.R. Strand},
keywords = {Many-body quantum physics, Imaginary time Green's functions, Low rank compression},
abstract = {We introduce libdlr, a library implementing the recently introduced discrete Lehmann representation (DLR) of imaginary time Green's functions. The DLR basis consists of a collection of exponentials chosen by the interpolative decomposition to ensure stable and efficient recovery of Green's functions from imaginary time or Matsubara frequency samples. The library provides subroutines to build the DLR basis and grids, and to carry out various standard operations. The simplicity of the DLR makes it straightforward to incorporate into existing codes as a replacement for less efficient representations of imaginary time Green's functions, and libdlr is intended to facilitate this process. libdlr is written in Fortran, provides a C header interface, and contains a Python module pydlr. We also introduce a stand-alone Julia implementation, Lehmann.jl.
Program summary
Program Title: libdlr CPC Library link to program files: https://doi.org/10.17632/56z594pzsj.1 Developer's repository link: https://github.com/jasonkaye/libdlr Licensing provisions: Apache-2.0 Programming language: Fortran, C, Python, Julia Nature of problem: Discretization and compression of functions (Green's functions and self-energies) with an imaginary time variable. Solution method: Explicit basis functions and discretization points obtained by low rank compression of the analytical continuation kernel.}
}

@Article{FFTW.jl-2005,
  author = {Frigo, Matteo and Johnson, Steven~G. },
  title = {The Design and Implementation of {FFTW3}},
  journal = {Proceedings of the IEEE},
  year = {2005},
  volume = {93},
  number = {2},
  pages = {216--231},
  doi = {10.1109/JPROC.2004.840301}
}

@phdthesis{blommelthesis,
  author       = {Thomas Blommel}, 
  title        = {Numerical Integration of the Kadanoff-Baym Equations},
  school       = {University of Michigan},
  year         = 2024,
}

@article{murray2024,
  title = {Nonequilibrium diagrammatic many-body simulations with quantics tensor trains},
  author = {Murray, Matthias and Shinaoka, Hiroshi and Werner, Philipp},
  journal = {Phys. Rev. B},
  volume = {109},
  issue = {16},
  pages = {165135},
  numpages = {12},
  year = {2024},
  month = {Apr},
  publisher = {American Physical Society},
  doi = {10.1103/PhysRevB.109.165135},
  url = {https://link.aps.org/doi/10.1103/PhysRevB.109.165135}
}

@article{delatorre2021,
  title = {Colloquium: Nonthermal pathways to ultrafast control in quantum materials},
  author = {de la Torre, Alberto and Kennes, Dante M. and Claassen, Martin and Gerber, Simon and McIver, James W. and Sentef, Michael A.},
  journal = {Rev. Mod. Phys.},
  volume = {93},
  issue = {4},
  pages = {041002},
  numpages = {31},
  year = {2021},
  month = {Oct},
  publisher = {American Physical Society},
  doi = {10.1103/RevModPhys.93.041002},
  url = {https://link.aps.org/doi/10.1103/RevModPhys.93.041002}
}

@article{murakami2023,
  title = {Photoinduced nonequilibrium states in Mott insulators},
  author = {Murakami, Yuta and Gole\ifmmode \check{z}\else \v{z}\fi{}, Denis and Eckstein, Martin and Werner, Philipp},
  journal = {Rev. Mod. Phys.},
  volume = {97},
  issue = {3},
  pages = {035001},
  numpages = {63},
  year = {2025},
  month = {Jul},
  publisher = {American Physical Society},
  doi = {10.1103/tkjh-lr83},
  url = {https://link.aps.org/doi/10.1103/tkjh-lr83}
}

@article{ZhangPronyAC,
  title = {Minimal pole representation and analytic continuation of matrix-valued correlation functions},
  author = {Zhang, Lei and Yu, Yang and Gull, Emanuel},
  journal = {Phys. Rev. B},
  volume = {110},
  issue = {23},
  pages = {235131},
  numpages = {13},
  year = {2024},
  month = {Dec},
  publisher = {American Physical Society},
  doi = {10.1103/PhysRevB.110.235131},
  url = {https://link.aps.org/doi/10.1103/PhysRevB.110.235131}
}

@article{joost2020,
  title = {G1-G2 scheme: Dramatic acceleration of nonequilibrium Green functions simulations within the Hartree-Fock generalized Kadanoff-Baym ansatz},
  author = {Joost, Jan-Philip and Schl\"unzen, Niclas and Bonitz, Michael},
  journal = {Phys. Rev. B},
  volume = {101},
  issue = {24},
  pages = {245101},
  numpages = {27},
  year = {2020},
  month = {Jun},
  publisher = {American Physical Society},
  doi = {10.1103/PhysRevB.101.245101},
  url = {https://link.aps.org/doi/10.1103/PhysRevB.101.245101}
}

@article{schlunzen2020,
  title = {Achieving the Scaling Limit for Nonequilibrium Green Functions Simulations},
  author = {Schl\"unzen, Niclas and Joost, Jan-Philip and Bonitz, Michael},
  journal = {Phys. Rev. Lett.},
  volume = {124},
  issue = {7},
  pages = {076601},
  numpages = {6},
  year = {2020},
  month = {Feb},
  publisher = {American Physical Society},
  doi = {10.1103/PhysRevLett.124.076601},
  url = {https://link.aps.org/doi/10.1103/PhysRevLett.124.076601}
}

@article{stahl2022,
  title = {Memory truncated Kadanoff-Baym equations},
  author = {Stahl, Christopher and Dasari, Nagamalleswararao and Li, Jiajun and Picano, Antonio and Werner, Philipp and Eckstein, Martin},
  journal = {Phys. Rev. B},
  volume = {105},
  issue = {11},
  pages = {115146},
  numpages = {13},
  year = {2022},
  month = {Mar},
  publisher = {American Physical Society},
  doi = {10.1103/PhysRevB.105.115146},
  url = {https://link.aps.org/doi/10.1103/PhysRevB.105.115146}
}

@article{picano2021,
  title = {Accelerated gap collapse in a Slater antiferromagnet},
  author = {Picano, Antonio and Eckstein, Martin},
  journal = {Phys. Rev. B},
  volume = {103},
  issue = {16},
  pages = {165118},
  numpages = {12},
  year = {2021},
  month = {Apr},
  publisher = {American Physical Society},
  doi = {10.1103/PhysRevB.103.165118},
  url = {https://link.aps.org/doi/10.1103/PhysRevB.103.165118}
}

@article{dasari2021,
  title = {Photoinduced strange metal with electron and hole quasiparticles},
  author = {Dasari, Nagamalleswararao and Li, Jiajun and Werner, Philipp and Eckstein, Martin},
  journal = {Phys. Rev. B},
  volume = {103},
  issue = {20},
  pages = {L201116},
  numpages = {6},
  year = {2021},
  month = {May},
  publisher = {American Physical Society},
  doi = {10.1103/PhysRevB.103.L201116},
  url = {https://link.aps.org/doi/10.1103/PhysRevB.103.L201116}
}

@Article{kaye2021,
	title={{Low rank compression in the numerical solution of the nonequilibrium {D}yson equation}},
	author={Jason Kaye and Denis Golež},
	journal={SciPost Phys.},
	volume={10},
	pages={091},
	year={2021},
	publisher={SciPost},
	doi={10.21468/SciPostPhys.10.4.091},
	url={https://scipost.org/10.21468/SciPostPhys.10.4.091},
}

@article{kaye22,
  title = {Discrete {L}ehmann representation of imaginary time {G}reen's functions},
  author = {Kaye, Jason and Chen, Kun and Parcollet, Olivier},
  journal = {Phys. Rev. B},
  volume = {105},
  issue = {23},
  pages = {235115},
  numpages = {18},
  year = {2022},
  month = {Jun},
  publisher = {American Physical Society},
  doi = {10.1103/PhysRevB.105.235115},
  url = {https://link.aps.org/doi/10.1103/PhysRevB.105.235115}
}

@article{shinaoka2023,
  title = {Multiscale Space-Time Ansatz for Correlation Functions of Quantum Systems Based on Quantics Tensor Trains},
  author = {Shinaoka, Hiroshi and Wallerberger, Markus and Murakami, Yuta and Nogaki, Kosuke and Sakurai, Rihito and Werner, Philipp and Kauch, Anna},
  journal = {Phys. Rev. X},
  volume = {13},
  issue = {2},
  pages = {021015},
  numpages = {27},
  year = {2023},
  month = {Apr},
  publisher = {American Physical Society},
  doi = {10.1103/PhysRevX.13.021015},
  url = {https://link.aps.org/doi/10.1103/PhysRevX.13.021015}
}

@article{Reeves2023,
  title = {Unimportance of memory for the time nonlocal components of the Kadanoff-Baym equations},
  author = {Reeves, Cian C. and Zhu, Yuanran and Yang, Chao and Vl\ifmmode \check{c}\else \v{c}\fi{}ek, Vojt\ifmmode \check{e}\else \v{e}\fi{}ch},
  journal = {Phys. Rev. B},
  volume = {108},
  issue = {11},
  pages = {115152},
  numpages = {15},
  year = {2023},
  month = {Sep},
  publisher = {American Physical Society},
  doi = {10.1103/PhysRevB.108.115152},
  url = {https://link.aps.org/doi/10.1103/PhysRevB.108.115152}
}

@article{karlsson2021,
  title = {Fast Green's Function Method for Ultrafast Electron-Boson Dynamics},
  author = {Karlsson, Daniel and van Leeuwen, Robert and Pavlyukh, Yaroslav and Perfetto, Enrico and Stefanucci, Gianluca},
  journal = {Phys. Rev. Lett.},
  volume = {127},
  issue = {3},
  pages = {036402},
  numpages = {8},
  year = {2021},
  month = {Jul},
  publisher = {American Physical Society},
  doi = {10.1103/PhysRevLett.127.036402},
  url = {https://link.aps.org/doi/10.1103/PhysRevLett.127.036402}
}

@misc{stefanucci2023,
      title={Semiconductor Electron-Phonon Equations: a Rung Above Boltzmann in the Many-Body Ladder}, 
      author={Gianluca Stefanucci and Enrico Perfetto},
      year={2023},
      eprint={2311.03980},
      archivePrefix={arXiv},
      primaryClass={cond-mat.mtrl-sci},
      url={https://arxiv.org/abs/2311.03980}, 
}

@article{kaye23,
author = {Kaye, Jason and Strand, Hugo U. R.},
title = {A Fast Time Domain Solver for the Equilibrium {D}yson Equation},
year = {2023},
_issue_date = {Aug 2023},
publisher = {Springer-Verlag},
address = {Berlin, Heidelberg},
volume = {49},
number = {4},
issn = {1019-7168},
_url = {https://doi.org/10.1007/s10444-023-10067-7},
doi = {10.1007/s10444-023-10067-7},
journal = {Adv. Comput. Math.},
_month = {aug},
numpages = {26}
}

@article{yin22,
title = {Using dynamic mode decomposition to predict the dynamics of a two-time non-equilibrium {G}reen’s function},
_journal = {Journal of Computational Science},
journal = {J. Comput. Sci.},
volume = {64},
pages = {101843},
year = {2022},
issn = {1877-7503},
doi = {https://doi.org/10.1016/j.jocs.2022.101843},
url = {https://www.sciencedirect.com/science/article/pii/S1877750322002022},
author = {Jia Yin and Yang-hao Chan and Felipe H. da Jornada and Diana Y. Qiu and Steven G. Louie and Chao Yang}
}

@article{yin23,
title = {Analyzing and predicting non-equilibrium many-body dynamics via dynamic mode decomposition},
_journal = {Journal of Computational Physics},
journal = {J. Comput. Phys.},
volume = {477},
pages = {111909},
year = {2023},
issn = {0021-9991},
doi = {https://doi.org/10.1016/j.jcp.2023.111909},
url = {https://www.sciencedirect.com/science/article/pii/S0021999123000049},
author = {Jia Yin and Yang-hao Chan and Felipe H. {da Jornada} and Diana Y. Qiu and Chao Yang and Steven G. Louie}
}

@article{reeves23,
  title = {Dynamic mode decomposition for extrapolating nonequilibrium {G}reen's-function dynamics},
  author = {Reeves, Cian C. and Yin, Jia and Zhu, Yuanran and Ibrahim, Khaled Z. and Yang, Chao and Vl\ifmmode \check{c}\else \v{c}\fi{}ek, Vojt\ifmmode \check{e}\else \v{e}\fi{}ch},
  journal = {Phys. Rev. B},
  volume = {107},
  issue = {7},
  pages = {075107},
  numpages = {10},
  year = {2023},
  month = {Feb},
  publisher = {American Physical Society},
  doi = {10.1103/PhysRevB.107.075107},
  url = {https://link.aps.org/doi/10.1103/PhysRevB.107.075107}
}

@unpublished{zhu24,
      title={Predicting nonequilibrium {G}reen's function dynamics and photoemission spectra via nonlinear integral operator learning}, 
      author={Yuanran Zhu and Jia Yin and Cian C. Reeves and Chao Yang and Vojtech Vlcek},
      year={2024},
      eprint={2407.09773},
      archivePrefix={arXiv},
      primaryClass={cond-mat.str-el},
      url={https://arxiv.org/abs/2407.09773}, 
}

@article{freericks2006,
  title = {Nonequilibrium Dynamical Mean-Field Theory},
  author = {Freericks, J. K. and Turkowski, V. M. and Zlati\ifmmode \acute{c}\else \'{c}\fi{}, V.},
  journal = {Phys. Rev. Lett.},
  volume = {97},
  issue = {26},
  pages = {266408},
  numpages = {4},
  year = {2006},
  month = {Dec},
  publisher = {American Physical Society},
  doi = {10.1103/PhysRevLett.97.266408},
  url = {https://link.aps.org/doi/10.1103/PhysRevLett.97.266408}
}

@article{lamic2024,
  title={Solving the Transient Dyson Equation with Quasilinear Complexity via Matrix Compression},
  author={Lamic, Baptiste},
  journal={arXiv preprint arXiv:2410.11057},
  year={2024}
}

@article{blommel25,
  title = {Chirped amplitude mode in photoexcited superconductors},
  author = {Blommel, Thomas and Kaye, Jason and Murakami, Yuta and Gull, Emanuel and Gole\ifmmode \check{z}\else \v{z}\fi{}, Denis},
  journal = {Phys. Rev. B},
  volume = {111},
  issue = {9},
  pages = {094502},
  numpages = {9},
  year = {2025},
  month = {Mar},
  publisher = {American Physical Society},
  doi = {10.1103/PhysRevB.111.094502},
  url = {https://link.aps.org/doi/10.1103/PhysRevB.111.094502}
}

@article{kovacevic2025,
  title = {Toward Numerically Exact Computation of Conductivity in the Thermodynamic Limit of Interacting Lattice Models},
  author = {Kova\ifmmode \check{c}\else \v{c}\fi{}evi\ifmmode \acute{c}\else \'{c}\fi{}, Jeremija and Ferrero, Michel and Vu\ifmmode \check{c}\else \v{c}\fi{}i\ifmmode \check{c}\else \v{c}\fi{}evi\ifmmode \acute{c}\else \'{c}\fi{}, Jak\ifmmode \check{s}\else \v{s}\fi{}a},
  journal = {Phys. Rev. Lett.},
  volume = {135},
  issue = {1},
  pages = {016502},
  numpages = {7},
  year = {2025},
  month = {Jul},
  publisher = {American Physical Society},
  doi = {10.1103/mm38-zttx},
  url = {https://link.aps.org/doi/10.1103/mm38-zttx}
}

@article{huang2019,
  title = {Strange metallicity in the doped Hubbard model},
  volume = {366},
  ISSN = {1095-9203},
  url = {http://dx.doi.org/10.1126/science.aau7063},
  DOI = {10.1126/science.aau7063},
  number = {6468},
  journal = {Science},
  publisher = {American Association for the Advancement of Science (AAAS)},
  author = {Huang,  Edwin W. and Sheppard,  Ryan and Moritz,  Brian and Devereaux,  Thomas P.},
  year = {2019},
  month = nov,
  pages = {987–990}
}

@article{vucicevic2019,
  title = {Conductivity in the Square Lattice Hubbard Model at High Temperatures: Importance of Vertex Corrections},
  author = {Vu\ifmmode \check{c}\else \v{c}\fi{}i\ifmmode \check{c}\else \v{c}\fi{}evi\ifmmode \acute{c}\else \'{c}\fi{}, J. and Kokalj, J. and \ifmmode \check{Z}\else \v{Z}\fi{}itko, R. and Wentzell, N. and Tanaskovi\ifmmode \acute{c}\else \'{c}\fi{}, D. and Mravlje, J.},
  journal = {Phys. Rev. Lett.},
  volume = {123},
  issue = {3},
  pages = {036601},
  numpages = {6},
  year = {2019},
  month = {Jul},
  publisher = {American Physical Society},
  doi = {10.1103/PhysRevLett.123.036601},
  url = {https://link.aps.org/doi/10.1103/PhysRevLett.123.036601}
}

@misc{inayoshi2025,
	title = {A causality-based divide-and-conquer algorithm for nonequilibrium {Green}'s function calculations with quantics tensor trains},
	url = {http://arxiv.org/abs/2509.15028},
	doi = {10.48550/arXiv.2509.15028},
	urldate = {2025-11-04},
	publisher = {arXiv},
	author = {Inayoshi, Ken and Środa, Maksymilian and Kauch, Anna and Werner, Philipp and Shinaoka, Hiroshi},
	month = sep,
	year = {2025},
	note = {arXiv:2509.15028 [cond-mat]},
}

@misc{sroda2025,
	title = {Predictor-corrector method based on dynamic mode decomposition for tensor-train nonequilibrium {Green}'s function calculations},
	url = {http://arxiv.org/abs/2509.22177},
	doi = {10.48550/arXiv.2509.22177},
	urldate = {2025-11-04},
	publisher = {arXiv},
	author = {Środa, Maksymilian and Inayoshi, Ken and Schüler, Michael and Shinaoka, Hiroshi and Werner, Philipp},
	month = sep,
	year = {2025},
	note = {arXiv:2509.22177 [cond-mat]},
}

@misc{gasperlin25,
      title={Stability and complexity of global iterative solvers for the {K}adanoff-{B}aym equations}, 
      author={Jo{\v z}e Ga{\v s}perlin and Denis Gole{\v z} and Jason Kaye},
      year={2025},
      eprint={2512.11371},
      archivePrefix={arXiv},
      primaryClass={cond-mat.str-el},
      url={https://arxiv.org/abs/2512.11371},
	  note={arXiv:2512.11371 [cond-mat]},
	  doi={10.48550/arXiv.2512.11371},
}

@article{Jarrell1996,
  title = {Bayesian inference and the analytic continuation of imaginary-time quantum Monte Carlo data},
  volume = {269},
  ISSN = {0370-1573},
  url = {http://dx.doi.org/10.1016/0370-1573(95)00074-7},
  DOI = {10.1016/0370-1573(95)00074-7},
  number = {3},
  journal = {Physics Reports},
  publisher = {Elsevier BV},
  author = {Jarrell,  Mark and Gubernatis,  J.E.},
  year = {1996},
  month = may,
  pages = {133–195}
}

@article{sandvik1998,
  title = {Stochastic method for analytic continuation of quantum Monte Carlo data},
  author = {Sandvik, Anders W.},
  journal = {Phys. Rev. B},
  volume = {57},
  issue = {17},
  pages = {10287--10290},
  numpages = {0},
  year = {1998},
  month = {May},
  publisher = {American Physical Society},
  doi = {10.1103/PhysRevB.57.10287},
  url = {https://link.aps.org/doi/10.1103/PhysRevB.57.10287}
}

@article{Yoshimi2019,
  title = {SpM: Sparse modeling tool for analytic continuation of imaginary-time Green’s function},
  volume = {244},
  ISSN = {0010-4655},
  url = {http://dx.doi.org/10.1016/j.cpc.2019.07.001},
  DOI = {10.1016/j.cpc.2019.07.001},
  journal = {Computer Physics Communications},
  publisher = {Elsevier BV},
  author = {Yoshimi,  Kazuyoshi and Otsuki,  Junya and Motoyama,  Yuichi and Ohzeki,  Masayuki and Shinaoka,  Hiroshi},
  year = {2019},
  month = nov,
  pages = {319–323}
}

@article{fournier2020,
  title = {Artificial Neural Network Approach to the Analytic Continuation Problem},
  author = {Fournier, Romain and Wang, Lei and Yazyev, Oleg V. and Wu, QuanSheng},
  journal = {Phys. Rev. Lett.},
  volume = {124},
  issue = {5},
  pages = {056401},
  numpages = {6},
  year = {2020},
  month = {Feb},
  publisher = {American Physical Society},
  doi = {10.1103/PhysRevLett.124.056401},
  url = {https://link.aps.org/doi/10.1103/PhysRevLett.124.056401}
}

@article{yoon2018,
  title = {Analytic continuation via domain knowledge free machine learning},
  author = {Yoon, Hongkee and Sim, Jae-Hoon and Han, Myung Joon},
  journal = {Phys. Rev. B},
  volume = {98},
  issue = {24},
  pages = {245101},
  numpages = {7},
  year = {2018},
  month = {Dec},
  publisher = {American Physical Society},
  doi = {10.1103/PhysRevB.98.245101},
  url = {https://link.aps.org/doi/10.1103/PhysRevB.98.245101}
}

@article{Ying2022,
  title = {Analytic continuation from limited noisy Matsubara data},
  volume = {469},
  ISSN = {0021-9991},
  url = {http://dx.doi.org/10.1016/j.jcp.2022.111549},
  DOI = {10.1016/j.jcp.2022.111549},
  journal = {Journal of Computational Physics},
  publisher = {Elsevier BV},
  author = {Ying,  Lexing},
  year = {2022},
  month = nov,
  pages = {111549}
}

@article{sentef2013,
  title = {Examining Electron-Boson Coupling Using Time-Resolved Spectroscopy},
  author = {Sentef, Michael and Kemper, Alexander F. and Moritz, Brian and Freericks, James K. and Shen, Zhi-Xun and Devereaux, Thomas P.},
  journal = {Phys. Rev. X},
  volume = {3},
  issue = {4},
  pages = {041033},
  numpages = {11},
  year = {2013},
  month = {Dec},
  publisher = {American Physical Society},
  doi = {10.1103/PhysRevX.3.041033},
  url = {https://link.aps.org/doi/10.1103/PhysRevX.3.041033}
}

@article{picano2025,
  title = {Quantum Thermalization via Travelling Waves},
  author = {Picano, Antonio and Biroli, Giulio and Schir\`o, Marco},
  journal = {Phys. Rev. Lett.},
  volume = {134},
  issue = {11},
  pages = {116503},
  numpages = {7},
  year = {2025},
  month = {Mar},
  publisher = {American Physical Society},
  doi = {10.1103/PhysRevLett.134.116503},
  url = {https://link.aps.org/doi/10.1103/PhysRevLett.134.116503}
}

\end{document}